\documentclass{article}
\usepackage{graphicx}

\usepackage{jheppub}
\usepackage{amsmath}
\usepackage{epsfig,multicol,bbm}
\usepackage{url}
\usepackage{graphicx}
\usepackage{setspace}
\usepackage{amssymb}
\usepackage{listings}
\usepackage{epstopdf}
\usepackage{float}
\usepackage{ulem}
\usepackage{color}
\usepackage{amsmath}
\usepackage{ dsfont }
\usepackage{slashed}
\title{Towards the Natural Gauge Mediation}

\date{\today}

\author[1]{Ran Ding}
\author[2,3]{Tianjun Li}
\author[4]{Liucheng Wang}
\author[2,5]{Bin Zhu}
\affiliation[1]{Center for High-Energy
Physics, Peking University, Beijing, 100871, P. R. China}
\affiliation[2]{State Key Laboratory of Theoretical Physics
and Kavli Institute for Theoretical Physics, China (KITPC), 
Institute of Theoretical Physics, Chinese Academy of Sciences,
Beijing 100190, P. R. China}
\affiliation[3]{School of Physical Electronics,
University of Electronic Science and Technology of China, 
Chengdu 610054, P. R. China}
\affiliation[4]{Bartol Research Institute, Department of Physics and Astronomy,
University of Delaware, Newark, DE 19716, USA}
\affiliation[5]{Institute of Physics Chinese Academy of sciences, Beijing 100190, P. R. China}

\emailAdd{dingran@mail.nankai.edu.cn}
\emailAdd{tli@itp.ac.cn}
\emailAdd{lcwang@udel.edu}
\emailAdd{zhubin@mail.nankai.edu.cn}

\abstract
{The sweet spot supersymmetry (SUSY) solves the $\mu$ problem in the Minimal Supersymmetric Standard Model (MSSM) with gauge mediated SUSY breaking (GMSB) via the generalized Giudice-Masiero (GM) mechanism where only the $\mu$-term and soft Higgs  masses are generated at the unification scale of the Grand Unified Theory (GUT) due to the
approximate PQ symmetry. Because all the other SUSY breaking soft terms are generated via the GMSB below the GUT scale, there exists SUSY electroweak (EW) fine-tuning problem to explain the 125~GeV Higgs boson mass due to small trilinear soft term. Thus, to explain the Higgs boson mass, we propose the GMSB with both the generalized GM mechanism and Higgs-messenger interactions. The renormalization group equations are runnings from the GUT scale down to EW scale. So the EW symmetry breaking can be realized easier. We can keep the gauge coupling unification and solution to the flavor problem in the GMSB, as well as solve the $\mu/B_{\mu}$-problem. Moreover, there are only five free parameters in our model. So we can determine the characteristic low energy spectra and explore its distinct phenomenology. The low-scale fine-tuning measure can be as low as 20 with the light stop mass below 1 TeV and gluino mass below 2 TeV. 
The gravitino dark matter can come from a thermal production with the correct relic density and be consistent with the thermal leptogenesis. Because gluino and stop can be relatively light in our model, how to search for such GMSB at the upcoming run II of the LHC experiment could be very interesting.}

\begin{document}
\maketitle

% =============================================================================
\section{Introduction}
\label{sec:intro}
% =============================================================================

A Higgs boson with mass around 125 GeV has been discovered at the LHC
by both ATLAS and CMS Collaborations~\cite{Aad:2012tfa,Chatrchyan:2012ufa}.
After the run I of the LHC, it had been proven to behave,
interact and decay in many of the ways similar to the Standard Model (SM) Higgs boson. 
More precision measurements are needed to determine if the discovered particle is
exactly the SM Higgs boson, or whether multiple Higgs bosons and exotic
decays exist as predicted by some other models. A SM-like Higgs boson
with mass around 125 GeV renews the hierarchy problem as the quadratic
divergences of the quantum corrections to its mass are a major concern
from the theoretical perspective. The electroweak-scale supersymmetry (SUSY) remains an
elegant solution to this problem and is still a promising extension
of the SM. A SM-like Higgs boson with mass 125 GeV can be identified
as the light CP-even Higgs boson $h$ in  the Minimal
Supersymmetric Standard Model (MSSM) (See, for example, \cite{Carena:1995wu,Martin:1997ns}.). 
If all the other Higgs bosons are heavy, the Higgs sector will fall into the decoupling MSSM limit, 
where the properties of $h$ are similar to the SM Higgs boson. 
The loop contributions to the Higgs mass $m_{h}$ have to be significant as
the tree-level $m_{h}$ is smaller than the $Z$ boson mass $M_Z$~\cite{Inoue:1982ej,Flores:1982pr}. 
Although the two-loop \cite{Goodsell:2014bna} and even three-loop
contributions \cite{Feng:2013tvd} are important to achieve the mass $m_{h}$ around 125
GeV,  general features can be determined by the dominating one-loop contributions
from top-stop sector as follows
\begin{equation}
m_{h}^{2}\simeq m_{Z}^{2}\cos^{2}2\beta+\frac{3m_{t}^{4}}{4\pi^{2}v^{2}}\left[\log\frac{M_{\mathrm{SUSY}}^{2}}{m_{t}^{2}}+\frac{\tilde{A}_{t}^{2}}{M_{\mathrm{SUSY}}^{2}}\left(1-\frac{\tilde{A}_{t}^{2}}{12M_{\mathrm{SUSY}}^{2}}\right)\right],\label{eq:higgs mass}
\end{equation}
where $m_{t}$ is the top quark mass, $v=174$ GeV is vacuum expectation value (VEV)
for electroweak
symmetry breaking (EWSB), $M_{\mathrm{SUSY}}=\sqrt{m_{\tilde{t}_{1}}m_{\tilde{t}_{2}}}$
is the geometric mean of stop masses, and $\tilde{A}_{t}$ is defined by 
\begin{equation}
\tilde{A}_{t}=A_{t}-\mu\cot\beta.\label{eq:At}
\end{equation}
Here $A_{t}$ is the trilinear soft term for Higgs-stop coupling, 
$\mu$ is the bilinear Higg boson mass in the MSSM superpotential, and
$\tan\beta=\langle H_u\rangle/\langle H_d\rangle$ is the ratio of two Higgs VEVs. 
One can choose $M_{\mathrm{SUSY}}^{2}/m_{t}^{2}\gg1$ in Eq (\ref{eq:higgs mass})
to enhance the loop contribution. The stop masses have to be larger
than 10 TeV if there is no stop mixing. This set of parameters will
result in a relatively heavy SUSY spectrum, which violates the 
naturalness condition and cannot have any meaningful stop signals at the LHC. 
Therefore, in this paper, we focus on another milder way to have a large
loop contribution by choosing $M_{\mathrm{SUSY}}^{2}/m_{t}^{2}>1$
and $\tilde{A}_{t}^{2}/M_{\mathrm{SUSY}}^{2}>1$ in Eq (\ref{eq:higgs mass}).
Namely, the geometric mean of stop masses is larger than 1 TeV as well as a large mixing
parameter $\tilde{A}_{t}$. The maximal mixing happens when $\tilde{A}_{t}\sim\sqrt{6}M_{\mathrm{SUSY}}$ \cite{Carena:2011aa}. However, such a maximal mixing scenario may
 lead to a color-breaking minimum where
the stops have non-vanishing VEVs \cite{Camargo-Molina:2013sta,Chowdhury:2013dka,Blinov:2013fta,MOLINA:2014uha,Camargo-Molina:2014pwa,Chattopadhyay:2014gfa,Camargo-Molina:2013sna}. 

Besides the discovery of the Higgs boson, no signals of SUSY particles have been observed
at the run I of the LHC. Although the compressed SUSY are always hard to be tested/excluded due to the cancellation of missing energy~\cite{LeCompte:2011fh,LeCompte:2011cn}, squarks and gluino are in general forced
to be heavy after the LHC8. Together with a 125 GeV Higgs
boson, it raises uncomfortable issues with naturalness widely discussed
in literatures. As we know, there are usually three kinds of ways to estimate
the SUSY breaking effects from the hidden sector into visible MSSM sector:
gravity, gauge, and anomaly mediations. In gravity mediation, the SUSY breaking soft
terms are generally obtained by the high-dimension operators suppressed by the reduced 
Planck scale $M_{\mathrm{PL}}$. A large $A_t$  can be obtained from a ultraviolet (UV) 
boundary condition or from the evolution of the
renormalization group  equations (RGEs) from $M_{\mathrm{PL}}$ to the electroweak (EW)
scale $M_{\text{EW}}$, which will significantly enhance the Higgs mass $m_h$. 
Because the gravity effects are universal to three generations, 
their soft masses and A-terms are not generation-blind. 
So gravity mediation always suffers from the flavor problem. 
In contract,  gauge mediation is flavor-safe as the corresponding operators of sfermions
are all aligned. But the challenges appear in the Higgs sector. 
In the gauge mediation SUSY breaking (GMSB), 
A-terms are vanishing at one-loop level when the messengers are integrated out. 
In order to get a sufficiently large $A_t$-term at the EW scale, we must
either have a heavy gluino in the model, or run the RGEs for a long scale range
by assuming high-scale SUSY breaking. Besides the necessary Higgs
mass corrections from a large $A_{t}$-term, it is also unclear how
to generate an appropriate size of $\mu$-term in the GMSB while keeping $B_{\mu}$ 
around the same scale (for a review of the $\mu$/$B_{\mu}$ problem, see \cite{Giudice:1998bp}.). 
The $\mu$-term is a bilinear Higgs mass in the superpotential
\begin{equation}
W \supset \mu H_u H_d.\label{eq:mu_definition}
\end{equation}
A successful EW symmetry breaking (EWSB) requires the $\mu$-term to be
 the same order of the SUSY breaking soft mass, namely $\mu\sim m_{\text{soft}}\ll M_{\text{PL}}$.
In the gravity mediation, an appropriate size of $\mu$-term can be obtained by 
the Giudice-Masiero mechanism \cite{Giudice:1988yz}.
However, the minimal GMSB does not generate the $\mu$-term when the messenger fields
are integrated out. In fact, the $\mu$-term can be forbidden if there exists a Peccei-Quinn (PQ) symmetry. 
An appropriate size of $\mu$-term can be obtained if the PQ symmetry is broken or just an approximate one.
In the GMSB, a simple way to break the PQ symmetry is adding Yukawa couplings between 
the Higgs sector and messengers in the superpotential. 
Hence the $\mu$-term can be naturally generated via one-loop Feynman diagrams 
at the messenger scale~\cite{Dvali:1996cu}. 
However, the correspoding soft term $B_{\mu}$ is generated at one-loop level 
as well 
\begin{equation}
\mathcal{L}_{\text{soft}}\supset B_{\mu} H_u H_d. 
\end{equation}
Thus, the $B_{\mu}$-term is too large compared to $\mu$-term squared 
by a loop factor, {\it i.e.}, $B_{\mu}\sim16\pi^{2}\mu^{2}$.
Since a successful EWSB requires $B_{\mu}\sim\mu^{2}$, this is the $\mu$/$B_{\mu}$ 
problem in the GMSB.
One simple solution is extending the MSSM to the
next to MSSM (NMSSM) \cite{Ellwanger:2009dp}, where a new SM
singlet is coupled to Higgs fields as well as messengers. 
The $\mu$/$B_{\mu}$ problem also exists in the anomaly mediation, 
where the couplings between the visible and hidden
sectors are much more suppressed than by the reduced Planck scale due to the one-loop
suppressions. 
In addition, the simple anomaly mediation further suffers the tachyonic problem 
as the slepton mass squared are predicted to be negative.

Since we are waiting for the run II of the LHC, it is important to think about the
feasible SUSY models to describe physics at the TeV scale. 
Although the naturalness assumption is challenged by the existing results of the LHC, 
no other serious paradigm has appeared to replace it.
So we still take the naturalness assumption as a guiding principle in 
constructing SUSY models. 
All mentioned problems should be addressed without moving forward into the relatively heavy SUSY spectra~\cite{Giudice:2004tc,ArkaniHamed:2004yi}. As we know, in the framework of the so-called sweet 
spot SUSY \cite{Ibe:2007km,Ibe:2007gf,Ibe:2007mr,Fukushima:2013vxa},
the SUSY breaking sector and  Higgs fields are directly coupled at 
the unification  scale $\Lambda_{\mathrm{GUT}}\sim10^{16}$ GeV in the Grand Unified Theories (GUTs). 
Because the whole sector respects the approximate PQ symmetry, $\mu$-term is generated at $\Lambda_{\mathrm{GUT}}$ scale
by the generalized Giudice-Masiero (GM) mechanism \cite{Giudice:1988yz} with a vanishing $B_{\mu}$-term.
Below $\Lambda_{\mathrm{GUT}}$ it is effectively the GMSB, and then 
the soft masses of SUSY particles are mainly obtained after the messenger fields 
are integrated out. There is generally no flavor problem since the gravitino mass $m_{3/2}$ is typically smaller than $\mathcal{O}(1)$ GeV. On the other hand, to generate a non-vanishing $A_t$-term at the messenger scale and lift the Higgs boson mass, we can introduce 
the Higgs-messengers interaction~\cite{Kang:2012ra,Craig:2012xp,Albaid:2012qk,Byakti:2013ti,Evans:2013kxa,Knapen:2013zla,Ding:2013pya,Ding:2014bqa}.
Therefore, in this paper, we shall propose the GMSB with 
 the generalized GM mechanism and Higgs-messenger interaction.
 Our model can have a SM-like Higgs boson with mass 125 GeV without moving forward 
into the relatively heavy SUSY spectra.
We also show that the current LHC SUSY search bounds can be evaded.
The low-scale fine-tuning measure can be as low as 20 in this model with 
the light stop mass below 1 TeV. Moreover,
 the gravitino is the lightest supersymmetric particle (LSP) and can be a good dark matter 
candidate which is consistent with the relic density observation via thermal production.
This natural SUSY scenario could be an interesting scenario at the coming run II of 
the LHC experiment as it is theoretically supported and simply predicted by only five parameters.

This paper is organized as follows. In Section II, we will consider the model in details.
Section III is devoted to studying the viable parameter spaces, which are
 consistent with all the current LHC observations and contain a good dark matter candidate. 
Finally, our conclusion is given in Section IV.

\section{The Natural GMSB}
\label{sec:model}

In this section, we present the GMSB with the generalized GM mechanism and 
Higgs-messenger interaction.
The discovery of a SM-like Higgs boson at 125 GeV as well as the natural SUSY assumption 
indicates a large $A_t$-term in the MSSM. In order to generate a non-vanishing $A_t$-term at the messenger scale, an extended Higgs-messenger coupling $\lambda_uH_u\Phi_1\bar{\Phi}_2$ has always been introduced in GMSB~\cite{Kang:2012ra,Craig:2012xp,Albaid:2012qk,Byakti:2013ti,Evans:2013kxa,Knapen:2013zla,Ding:2013pya,Ding:2014bqa}.
In those SUSY models, the Yukawa coupling $\lambda_d$ between $H_d$ and messenger fields always turns off, otherwise the $\mu/B_{\mu}$-problem will show up. In order to obtain an appropriate $\mu$-term in our model, 
we assume that the SUSY breaking sector and the Higgs fields are directly coupled at the GUT scale 
$\Lambda_{\mathrm{GUT}}$, as in the sweet spot SUSY \cite{Ibe:2007km,Ibe:2007gf,Ibe:2007mr,Fukushima:2013vxa}. 
Because of the approximate PQ symmetry, only the $\mu$-term and soft masses $m_{H_u}$/$m_{H_d}$ are 
generated at $\Lambda_{\mathrm{GUT}}$. The sfermion soft masses, gaugino soft masses, $A$-terms,
 and $B_{\mu}$-term are all vanished at $\Lambda_{\mathrm{GUT}}$.
Below $\Lambda_{\mathrm{GUT}}$ it is effectively the GMSB with extended Higgs-messenger coupling.
The RGEs are runnings from the GUT scale to the EW scale. 
At the messenger scale, the messenger fields should be integrated out,
and the non-vanishing soft masses of the gauginos/sfermions and A-terms are generated as 
threshold corrections in the RGEs. 
Such effects from the gravity mediation are tiny as the gravitino mass $m_{3/2}$ is assumed to be typically smaller 
than $\mathcal{O}(1)$ GeV.
In this model, the gauge coupling unification is guaranteed. The flavor problem and $\mu/B_{\mu}$-problem are solved.

\subsection{Supersymmetry Breaking}
A consequence of SUSY spontaneously breaking is the existence of a massless Goldstone
fermion, the Goldstino. For a F-term SUSY breaking theory, one always
assumes a chiral singlet superfield $X$, which is formed by the Goldstino,
its superpartner sGoldstino, and its non-vanishing F-term. A broad class
of SUSY breaking models can be described by the Polonyi model as a
low-energy effective theory. The Polonyi model is given by the corresponding
K\"ahler potential and superpotential as 
\begin{equation}
\mathcal{L}=\int d^{4}\theta\left[X^{\dagger}X-\frac{\left(X^{\dagger}X\right)^{2}}{\Lambda_X^{2}}\right]+\left[\int d^{2}\theta fX+\mathrm{h.c.}\right].\label{eq:X}
\end{equation}
Here $\Lambda_X$ is the typical mass scale where the heavy particles
have been integrated out. This effective description is valid as long
as $f<\Lambda_X^{2}$ and can be realized in many UV completed
models, for example, the O'Raifeartaigh model \cite{O'Raifeartaigh:1975pr} 
and SUSY QCD models with a meta-stable vacuum \cite{Intriligator:2006dd}. 
The chiral superfield $X$ can even be a composite filed if the UV completed models are some strongly coupled gauge theories \cite{Izawa:1996pk, Intriligator:1996pu}.
Based on Eq.~(\ref{eq:X}), $F_{X}=-f\neq0$ is obtained by the equation of motion.
The positive energy of the vacuum breaks SUSY spontaneously and $X=0$ is the position of vacuum of the potential. 

In the gauge mediation, the vector-like messenger superfields $\Phi$ and $\bar{\Phi}$ will couple to the SUSY breaking sector generally via a superpotential $W=\kappa X \Phi \bar{\Phi}$. 
However, the F-component of $X$ in this case is $F_{X}=-f-\kappa \Phi \bar{\Phi}$, which will lead to a SUSY-conserving minimum with $X=0$ and $\Phi \bar{\Phi}=-f/\kappa$.
In other words, SUSY will be restored after the naive introduction of the messenger fields coupling to 
the SUSY breaking sector.
Several baroque mechanisms have been discussed in order to guarantee a SUSY-breaking meta-stable vacuum 
in the gauge mediation~\cite{Shirman:1996jx,ArkaniHamed:1997ut,Murayama:1997pb,Izawa:1997gs,ArkaniHamed:1997jv,Kitano:2006wz}. 
For example, a SUSY-breaking vaccum away from the origin $X=0$ can be realized after taking the supergravity effect into account \cite{Kitano:2006wz}.
The minimum is at $X\sim \Lambda_X^2/M_{\mathrm{PL}}$ with $F_X\neq0$.
So a spurion structure $X=\langle X \rangle+F_X\theta^2$ can be assumed to parameterize the typical effects of SUSY breaking. 
It is important to have a SUSY-breaking vacuum away from the origin as the messenger mass $\kappa\langle X \rangle$ originally comes from the superpotential $W=\kappa X \Phi \bar{\Phi}$. 

\subsection{$\mu$-Term in Sweet Spot SUSY}

A successful EWSB puts two constraints at the EW scale on the Higgs sector of the MSSM 
including the $\mu$-term, which are shown as follows
\begin{align}
\sin2\beta&=\frac{2B_{\mu}}{2\mu^2+m^2_{H_u}+m^2_{H_d}},\label{eq:EWSB1}\\
\frac{m^2_Z}{2}&=\frac{m^2_{H_d}-m^2_{H_u}\tan^2\beta}{\tan^2\beta-1}-\mu^2,\label{eq:EWSB2}
\end{align}
where $m^2_{H_u}$ and $m^2_{H_d}$ are the soft masses of $H_u$ and $H_d$, respectively. 
From Eq. (\ref{eq:EWSB1}) we know that $B_{\mu}\sim \mu^2$ at the EW scale. Moreover, 
for a moderately large $\tan\beta$
Eq. (\ref{eq:EWSB2}) can be simplified as below
\begin{equation}
m_{Z}^{2}\approx-2\left(\mu^{2}+m_{H_{u}}^{2}\right)~.~
\end{equation}
Here $m_{H_{u}}^{2}$ should be negative at the electroweak scale, which is
required by the EWSB. 
A natural EWSB requires that the cancellation between $\mu^2$ and $m^2_{H_{u}}$ be relatively small. 
Namely, it is unnatural that $\mu$-term is much larger than $m_{Z}$ at the electroweak scale 
although it is supersymmetric.
The scale of $\mu$ coincides with the soft mass.
This is the so-called $\mu$-problem: how to generate such an appropriate $\mu$-term in SUSY models.
Because of $\mu \ll M_{\mathrm{PL}}$, one can always assume that the $\mu$-term is prohibited by some symmetry 
and induced by a small breaking of such a symmetry. 
The requirement $B_{\mu}\sim \mu^2$ at the electroweak scale always results 
in the so-called $B_{\mu}$-problem in the GMSB, if it cannot be satisfied.

No matter how the SUSY breaking effects translate into the MSSM Higgs sector,
an effective K\"ahler potential between the SUSY breaking sector $X$ and Higgs sector 
can be obtained as follows
\begin{equation}
\mathcal{K}_{\mathrm{eff}}=Z_{H_{u}}(X,\, X^{\dagger})H_{u}^{\dagger}H_{u}+Z_{H_{d}}(X,\, X^{\dagger})H_{d}^{\dagger}H_{d}+\left[Z_{H_{u}H_{d}}(X,\, X^{\dagger})H_{u}H_{d}+\mathrm{h.c.}\right]+...~.
\end{equation}
Here all the wavefunctions depend on some dimensional scale and
can be determined from a specific UV completed theory. We expand all the wavefunctions 
\begin{align}
\begin{cases}
Z_{H_{u}}(X,\, X^{\dagger})& =1+(a_{1}X+a_{1}^{*}X^{\dagger})+a_{2}X^{\dagger}X+...,\\
Z_{H_{d}}(X,\, X^{\dagger})& =1+(b_{1}X+b_{1}^{*}X^{\dagger})+b_{2}X^{\dagger}X+...,\\
Z_{H_{u}H_{d}}(X,\, X^{\dagger})& =c_{0}+(c_{1}X+c_{1}^{*}X^{\dagger})+c_{2}X^{\dagger}X+...,
\end{cases}\label{eq:wavefunction}
\end{align}
where both $Z_{H_u}$ and $Z_{H_d}$ are canonically normalized. These terms are responsible for generating $A_{u}$, $m_{H_{u}}^{2}$,
$A_{d}$, $m_{H_{d}}^{2}$, $\mu$ and $B_{\mu}$. To the leading order,
\begin{align}
\begin{cases}
A_{\mu} & =F_{X}\frac{\partial Z_{H_{u}}}{\partial X},\\
m_{H_u}^{2} & =F_{X}^{\dagger}F_{X}\frac{\partial^{2}Z_{H_{u}}}{\partial X^{\dagger}\partial X},\\
A_{d} & =F_{X}\frac{\partial Z_{H_{d}}}{\partial X},\\
m_{H_d}^{2} & =F_{X}^{\dagger}F_{X}\frac{\partial^{2}Z_{H_{d}}}{\partial X^{\dagger}\partial X},\\
\mu & =F_{X}^{\dagger}\frac{\partial Z_{H_{u}H_{d}}}{\partial X},\\
B_{\mu} & =F_{X}^{\dagger}F_{X}\frac{\partial^{2}Z_{H_{u}H_{d}}}{\partial X^{\dagger}\partial X}.
\end{cases}\label{eq:mu}
\end{align}
In supergravity, all the coefficients $a_i$, $b_j$, and $c_k$ in Eq. (\ref{eq:wavefunction}) are suppressed by $M_{\mathrm{PL}}$. In the unit of $M_{\mathrm{PL}}=1$, all the coefficients are actually
$\mathcal{O}(1)$. This is the Giudice-Masiero mechanism \cite{Giudice:1988yz}, which will
lead to the desired relation $\mu^{2}\sim B_{\mu}\sim m^2_{\mathrm{soft}}\ll M^2_{\text{Pl}}$. 
Unfortunately, gravity mediation always suffers from the flavor problem as the gravity effects 
are universal to three generations. In the GMSB, the $\mu$-term can be generated by adding couplings 
between the Higgs sector and messengers in the superpotential. 
Hence $\mu^{2}\sim m^2_{\mathrm{soft}}$ can be naturally achieved since all are generated at one-loop level. 
However, the $B_{\mu}$-term is also generated at one loop. 
This implies that $B_{\mu}$-term is too large by a loop factor compared to $\mu$-term squared 
as $B_{\mu}\sim 16\pi^{2}\mu^{2}$.
This is the $\mu/B_{\mu}$-problem in the gauge mediation.
An analogous problem, the $A/m^2_H$ problem in the gauge mediation, 
draws a lot of attention after the discovery of the 125 GeV Higgs boson \cite{Craig:2012xp}. 
In the gauge mediation, both $A$-term and the soft mass $m^2_H$ can be generated at the same loop order.
Since a large $A_t$-term is preferred by the Higgs discovery as well as the natural SUSY assumption, 
the corresponding large $m^2_H$ will seriously affect the EWSB, {\it i.e.},
the EWSB may not be realized.

In this paper, we base on the framework of the so-called sweet spot SUSY \cite{Ibe:2007km,Ibe:2007gf,Ibe:2007mr,Fukushima:2013vxa} to solve the $\mu/B_{\mu}$-problem.
Sweet spot SUSY is a phenomenological effective Lagrangian with certain natural assumptions,
which is designed to avoid problems in low energy phenomenology.
In this framework, the SUSY breaking sector and the Higgs fields are assumed to 
be directly coupled at the some energy scale. 
The PQ charge to $H_u$, $H_d$ and $X$ are assigned as follows
\begin{equation}
\mathrm{PQ}(H_u)=1,\: \mathrm{PQ}(H_d)=1, \:\mathrm{PQ}(X)=2.
\end{equation}
Then the wavefunctions in Eq. (\ref{eq:wavefunction}) will be constrained due to such a PQ symmetry. At the learding order, we have
\begin{align}
\begin{cases}
Z_{H_{u}}(X,\, X^{\dagger})=1+c_{H_u}\frac{X^{\dagger}X}{\Lambda_{H}^{2}},\\ Z_{H_{d}}(X,\, X^{\dagger})=1+c_{H_d}\frac{X^{\dagger}X}{\Lambda_{H}^{2}},\\ 
Z_{H_{u}H_{d}}(X,\, X^{\dagger})=c_{\mu}\frac{X^{\dagger}}{\Lambda_{H}}. \label{eq:Lambda_H}
\end{cases}
\end{align}
Here $\Lambda_{H}$ is the energy scale where the Higgs fields are directly coupled to the hidden sector. 
Because of the PQ symmetry, only the $\mu$-term and the soft masses $m_{H_u}$, $m_{H_d}$ are generated 
at $\Lambda_{H}$.
The $B_{\mu}$-term is vanishing at the scale $\Lambda_{H}$ as the UV boundary condition and can be 
non-vanishing at the EW scale due to the RGE running. So the $\mu$-term is generated without 
 $B_{\mu}$-problem.
The PQ symmetry is approximate because it is explicitly breaking in the SUSY-breaking sector 
by the superpotential $W=fX$ in Eq. (\ref{eq:X}). The MSSM Higgs sector will receive the explicit 
and small breaking of this approximate PQ symmetry when it is directly coupled to the hidden sector 
below the energy scale $\Lambda_{H}$. 

$\Lambda_{H}$ is not necessary to be the exact hidden sector scale $\Lambda_X$ in Eq.~(\ref{eq:X}). 
However, there is a sweet spot in SUSY models with $\Lambda_{H}=\Lambda_X=\Lambda_{\mathrm{GUT}}\sim10^{16}$ GeV \cite{Ibe:2007km,Ibe:2007gf,Ibe:2007mr,Fukushima:2013vxa}, 
in which the gauge coupling unification is realized. 
%Moreover, the correct size of $\mu$-term will be guaranteed if we choose %$\Lambda_{H}=\Lambda_{\mathrm{GUT}}$.
%Since $m_{3/2}\sim\frac{F_X}{M_{\mathrm{PL}}}$ is typically around $\mathcal{O}(1)$ GeV in order to avoid flavor constraints,
%$\mu\sim\frac{F_X}{M_{H}}$ will automatically be around $\mathcal{O}(100)$ GeV, as required by the natural SUSY.
Though sweet spot SUSY is a phenomenological effective Lagrangian, the 
UV completed models can be realized in several ways \cite{Ibe:2007km,Ibe:2007gf}.
So the $\mu$-term and soft masses $m_{H_u}$/$m_{H_d}$ are generated at $\Lambda_{\mathrm{GUT}}$
while the sfermion soft masses, gaugino  masses, $A$-terms, and $B_{\mu}$-term are all vanishing.
This is the UV boundary conditions at $\Lambda_{\mathrm{GUT}}$ in our model as
\begin{align}
\mu(M_{\text{GUT}})&=c_{\mu}\frac{F^\dagger_X}{\Lambda_{H}},\\
m^2_{H_u}(M_{\text{GUT}})&=c_{H_u}\frac{F^\dagger_XF_X}{\Lambda_{H}^{2}},\\
m^2_{H_d}(M_{\text{GUT}})&=c_{H_d}\frac{F^\dagger_XF_X}{\Lambda_{H}^{2}},\\
B_{\mu}(M_{\text{GUT}})&=0,\\
M_{1,2,3}(M_{\text{GUT}})&=0,\\
m_{\tilde\phi}^2(M_{\text{GUT}})&=0,\\
A_{Y_{u,d,e}}(M_{\text{GUT}})&=0.
\end{align}

In the exact sweet spot SUSY models \cite{Ibe:2007km,Ibe:2007gf,Ibe:2007mr,Fukushima:2013vxa}, it is effectively the GMSB below $\Lambda_{\mathrm{GUT}}$ as
\begin{equation}
W_{\text{GMSB}}= \kappa X\Phi_{i}\bar\Phi_{i},
\label{eq:mediation}
\end{equation}
where the fields $\Phi_{i}$ and $\bar\Phi_{i}$ form the $5 \oplus \bar 5$ or $10 \oplus \bar {10}$ representation of SU(5) as the gauge coupling unification is preserved. 
The RGE runnings from $\Lambda_{\mathrm{GUT}}$ down to the messenger scale $M_{\text{mess}}$ will lead to 
the non-vanishing sfermion soft masses and a small correction to $\mu$-term.
At the messenger scale, the messenger fields will be integrated out, which will generate 
the non-vanishing soft masses of the gauginos and sfermions as threshold corrections.
This procedure called ``matching'' is another part of the boundary conditions of the exact sweet spot SUSY models. 
The MSSM spectra will be generated after running RGEs from the messenger scale to EW scale. 
However, as already mentioned in Ref.~\cite{Fukushima:2013vxa}, the exact sweet spot SUSY would result 
in a heavy spectrum in order to obtain a $125$ GeV Higgs boson.
In particular, the gluino mass must be around $5$ TeV as well as $M_{\mathrm{SUSY}}\sim 5$ TeV, 
which definitely raises the SUSY EW fine-tuning problem. 
Although the LHC is a QCD machine, the colored particles in this scenario are too heavy to 
be detected at the LHC experiments.
An solution to the heavy spectrum problem can be found in
Refs.~\cite{Kang:2012ra,Craig:2012xp,Albaid:2012qk,Byakti:2013ti,Evans:2013kxa,Knapen:2013zla,Ding:2013pya,Ding:2014bqa} by adding extra Higgs-messenger Yukawa coupling.
In this paper, we would like to add such couplings in the sweet spot SUSY, where the $\mu$-problem 
and the flavor problem are still evaded.
As the SUSY particles will become relatively light in the modified sweet spot SUSY, 
it is hopeful to test this scenario by the coming run II of the LHC.

\subsection{The GMSB with Higgs-Messenger Coupling}
 
The GMSB models can be extended by introducing new Yukawa couplings between the Higgs sector and messengers \cite{Kang:2012ra,Craig:2012xp,Albaid:2012qk,Byakti:2013ti,Evans:2013kxa,Knapen:2013zla,Ding:2013pya,Ding:2014bqa}. 
In this paper, we modestly modify the exact sweet spot SUSY models 
by including the a direct interaction between Higgs field $H_u$ and messengers $\Phi_1$, $\Phi_2$ as
\begin{equation}
\delta W_{\text{Extended GMSB}}=\lambda_uH_u\Phi_1\bar{\Phi}_2. \label{eq:Higgs}
\end{equation}
Due to the new coupling $\lambda_u$, the trilinear soft terms get the non-vanishing
 contributions $A_u\propto -\frac{\lambda_u^2\Lambda}{16\pi^2}$ at the messenger scale with $\Lambda=F_{X}/M_{\text{mess}}$. The RGE runnings will result in large A-terms at the EW scale, which are preferred by 
the Higgs discovery as well as the natural SUSY condition.
There is no extra flavor problems caused by the extended Higgs-messenger coupling. 
There must exist another symmetry between $H_u$ and $H_d$, 
otherwise we should have another Yukawa coupling $\lambda_d$ between $H_d$ and messenger fields. If both $\lambda_u$ and $\lambda_d$ are non-vanishing, the 
extra contributions to $\delta\mu$ and $\delta B_{\mu}$ are naturally generated at one loop at the messenger scale.
The dangerous $\mu$/$B_{\mu}$ problem could emerge again. In this paper, we turn off the coupling $\lambda_d$, which can be forbidden by introducing another symmetry between $H_u$ and $H_d$.

Now we can embed the MSSM into the modified sweet spot SUSY and assume that the effective model below $M_{\text{GUT}}$ reduces to the GMSB with an extended Higgs-messenger coupling $\lambda_u$. 
After the messenger fields are integrated out, the non-vanishing soft masses of the gauginos/sfermions and A-terms are generated at the messenger scale. In order to get the Higgs boson $m_h$ around 125 GeV, $\lambda_u$ is usually required to be quite large at the messenger scale like $\lambda_u\sim 1$. If the messenger fields form the $5 \oplus \bar 5$ representation of SU(5), the one-loop RGE running of $\lambda_u$ is dominated by $\lambda_u$ and $y_t$. Typically $\lambda_u$ reaches a Landau pole before $M_{\text{GUT}}$, which is particularly troublesome \cite{Craig:2012xp}. In contrast, $\lambda_u$ will not meet a Landau pole if the messengers form the $10 \oplus \bar10$ representation of SU(5). In $10 \oplus \bar10$ models, the RG evolution of $\lambda_u$ is given as
\begin{equation}
\beta_{\lambda_u}=\frac{\lambda_u}{16\pi^2}\left[(3n_{10}+3)\lambda_u^2+3y_t^2-\frac{16}{3}g^2_3+...\right].
\end{equation}
The large negative contributions from $g_3$ would help to control the running of $\lambda_u$. In the paper, we choose the messenger fields as $10 \oplus \bar10$ models in order to evade a potential Landau pole problem of $\lambda_u$. 
Accordingly, $\Phi_1$ in Eq.~(\ref{eq:Higgs}) is in the (3,2,1/6) representation of the $10 \oplus \bar10$ messenger fields while $\Phi_2$ is in the (3,1,2/3) representation. The threshold corrections at the messenger scale $M_{\text{mess}}$ are given as
\begin{align}
\delta M_{a}(M_{\text{mess}})&= 3n_{10} \Lambda \frac{g_a^2(M_{\text{mess}})}{16\pi^2}g\left(\frac{\Lambda}{M_{\text{mess}}}\right)\,\,\,\,(a=1,2,3),\label{eq:mess1}\\
\delta m^2_{\tilde\phi}(M_{\text{mess}})&=3n_{10} \Lambda^2\sum_a C_a(k)\frac{g_a^4(M_{\text{mess}})}{(16\pi^2)^2}f\left(\frac{\Lambda}{M_{\text{mess}}}\right),\label{eq:mess2}\\
\delta A_{Y_{d,e}}(M_{\text{mess}})&=0,\label{eq:mess3}\\
\delta A_{Y_u}(M_{\text{mess}})&=-3n_{10}\Lambda\frac{\lambda_u^2}{16\pi^2}, \label{eq:mess4} \\
\delta m_{Q}^2(M_{\text{mess}})&=-3n_{10}\Lambda^2\frac{\lambda_u^2 y_t^2}{256\pi^4},\label{eq:mess5}\\
\delta m_{u}^2(M_{\text{mess}})&=-3n_{10}\Lambda^2\frac{\lambda_u^2 y_t^2}{128\pi^4},\label{eq:mess6}\\
\delta m_{H_u}^2(M_{\text{mess}})&=3n_{10}\Lambda^{2}\frac{(3+3n_{10})\lambda_{u}^{4}-2\sum_a C_a(k)g_{a}^{2}\lambda_{u}^{2}}{256\pi^{4}},\label{eq:mess7}
\end{align}
where we introduce $\Lambda=F_{X}/M_{\text{mess}}$. The first three equations 
(Eqs. (\ref{eq:mess1}), (\ref{eq:mess2}) and (\ref{eq:mess3})) are soft SUSY-breaking parameters 
in the original GMSB 
while the last four equations (Eqs. (\ref{eq:mess4}), (\ref{eq:mess5}), (\ref{eq:mess6}),
 and (\ref{eq:mess7})) are generated due to the extended Higgs-messenger coupling $\lambda_u$ 
 in Eq.~(\ref{eq:Higgs}). 
If we turn off the coupling $\lambda_u$ in Eq. (\ref{eq:Higgs}), the  threshold corrections 
shown in the last four equations will vanish.

It is easy to find out that our model depends on the following parameters
\begin{align}
\{\Lambda, M_{\text{mess}},\tan\beta,\lambda_u,n_{10}\}\oplus\{\mu(M_{\text{GUT}}),m^2_{H_u}(M_{\text{GUT}}),m^2_{H_d}(M_{\text{GUT}}),\alpha_{\text{GUT}},M_{\text{GUT}},Y_u,Y_d,Y_e\},
\label{eq:para1}
\end{align}
where $\alpha_{\text{GUT}}=g_{\text{GUT}}^2/4\pi$ with $g_{\text{GUT}}$  the unified gauge coupling constant.
 The parameter $\alpha_{\text{GUT}}$ is evaluated consistently with the experimental values of 
the electromagnetic constant $\alpha_{\text{em}}$, strong fine-structure constant $\alpha_{s}$, 
and the Weinberg angle $\sin^2\theta_{W}$ by solving RGEs numerically. The same integration procedure 
can also be applied to the Yukawa coupling constants $Y_u,~Y_d$, and $Y_e$. 
Therefore, the free parameters in Eq.~(\ref{eq:para1}) can be reduced to
\begin{align}
\{\Lambda, M_{\text{mess}},\tan\beta,\lambda_u, n_{10}\}\oplus\{\mu(M_{\text{GUT}}),m^2_{H_u}(M_{\text{GUT}}),m^2_{H_d}(M_{\text{GUT}})\}.
\label{eq:para2}
\end{align}
We emphasize that the soft masses of $H_u$ and $H_d$ are generated not only at the GUT scale but also 
at the messenger scale $M_{\text{mess}}$. 
Because the radiative EWSB is reproduced through the RGE effects on $m_{H_u}^2$, we can express
 $m_{H_u}^2$ and $m_{H_d}^2$ at the EW scale in terms of the 
other input parameters by minimizing the tree-level scalar potential 
\begin{align}
m_{H_u}^2&=-\mu^2+\frac{1}{2}M_{Z}^2\cos(2\beta)+B_{\mu}\cot\beta,\\
m_{H_d}^2&=-\mu^2-\frac{1}{2}M_{Z}^2\cos(2\beta)+B_{\mu}\tan\beta.
\end{align}
Thus, $m^2_{H_u}(M_{\text{GUT}})$ and $m^2_{H_d}(M_{\text{GUT}})$ are not free parameters, which
 are constrained by the successful EWSB. 
Of course, we should require $m^2_{H_u}(M_{\text{GUT}})>0$ and $m^2_{H_d}(M_{\text{GUT}})>0$
if the corresponding operators in the K\"ahler potential are generated at one loop.
In short, the free parameters of our model can be further reduced to
\begin{align}
\{\Lambda, M_{\text{mess}},\tan\beta,\lambda_u,n_{10}\}\oplus\{\mu(M_{\text{GUT}})\}.
\label{eq:para3}
\end{align}
We define $\mu(M_{\text{GUT}})=\mu_0$, which is the only free parameter at the GUT scale. 
Without losing the generality, we fix $n_{10}=1$ in this paper when we scan the parameter space. So finally, this model depends on only five free parameters 
\begin{align}
\{\Lambda, M_{\text{mess}},\tan\beta,\lambda_u,\mu_0 \}.
\label{eq:para4}
\end{align}
The $B_{\mu}$-term at the GUT scale vanishes automatically due to the approximate PQ symmetry,
 which is  one of our UV boundary conditions as well. 

We summarize our model here. 
At the GUT scale $\Lambda_{\mathrm{GUT}}$, the $\mu$-term and soft masses $m_{H_u}$/$m_{H_d}$ are 
generated as the visible Higgs sector receives the SUSY-breaking effects in Eq.~(\ref{eq:Lambda_H}). 
Only the parameter $\mu_0$ is a free parameter by requring the correct EWSB,
and the $B_{\mu}$-term vanishes at the GUT scale due to the PQ symmetry. 
Of course, we should require $m^2_{H_u}(M_{\text{GUT}})>0$ and $m^2_{H_d}(M_{\text{GUT}})>0$. 
Below $\Lambda_{\mathrm{GUT}}$ it is effectively the GMSB with an extended Higgs-messenger coupling, 
which is governed by the free parameters $\Lambda$, $M_{\text{mess}}$, and $\lambda_u$. 
At the messenger scale, the non-vanishing soft masses of the gauginos/sfermions and A-terms 
are generated as the threshold corrections, which are shown in Eqs. (\ref{eq:mess1} - \ref{eq:mess7}).
The effects from gravity mediation are negligible in our model as the gravity mass $m_{3/2}$ 
is assumed to be not larger than $\mathcal{O}(1)$ GeV. Therefore,
we construct a complete model in which a 125 SM-like Higgs boson is predicted, the flavor changing neutral currents are suppressed 
due to the gauge mediation, and the $\mu/B_{\mu}$ problem is naturally solved with 
the minimal set of parameters.

It is worth mentioning that the large trilinear $A_t$-term generated by the extended Higgs-messenger coupling $\lambda_u$ plays a crucial role in lifting the Higgs mass while keeping the MSSM spectrum light~\cite{Draper:2011aa}.
As a result, the fine-tuning in such kind of models generally becomes smaller compared to the conventional GMSB. 
However, the integration over the RGEs is not straightforward running
 from the GUT scale $M_{\text{GUT}}$ to SUSY scale $M_{\text{SUSY}}$. 
For the messenger thresholds, the additional soft terms are generated as shown in Eqs.~(\ref{eq:mess1} - \ref{eq:mess7}). The two-steps integration makes the high-scale fine-tuning parameters ill-defined in our model.
For example, we cannot use the hign-scale fine-tuning measures defined in Refs.~\cite{Ellis:1986yg, Barbieri:1987fn}.
Therefore, we will consider the low-scale fine-tuning measure.
In order to provide the possible quantitive measure of fine-tuning,  we employ
 the low-scale fine-tuning measure $\Delta_{\text{FT}}$ proposed in Refs.~\cite{Baer:2012mv,Baer:2013gva}
as follows
\begin{align}
C_{\mu}&=\left|\mu^2\right|,\nonumber\\
C_{B_{\mu}}&=\left|B_{\mu}\right|,\nonumber\\
C_{H_u}&=\left|\frac{m^2_{H_u}\tan^2\beta}{\tan^2\beta-1}\right|,\label{eq:FT} \\
C_{H_d}&=\left|\frac{m^2_{H_d}}{\tan^2\beta-1}\right|,\nonumber\\
\Delta_{\text{FT}}&=\frac{2}{M_Z^2}\text{max}(C_{\mu},C_{B_{\mu}},C_{H_u},C_{H_d}).\nonumber
\end{align}
In the next Section, we will present the detailed discussions about the MSSM spectra and 
phenomenological consequences.

\section{Numerical Results}
\label{sec:numerical}

We shall present the numerical studies of our model, 
including the particle spectra and low-scale fine-tuning measures. 
For this purpose, we implement this model in the Mathamatica package 
 {\tt SARAH}~\cite{Staub:2008uz,Staub:2009bi,Staub:2010jh,Staub:2012pb,Staub:2013tta} 
and generate the corresponding {\tt SPheno} file \cite{Porod:2011nf,Porod:2003um} to 
calculate the corresponding particle spectra. 
There are a lot of constraints on parameter spaces from the run I of the LHC. 
First, a SM-like Higgs boson at 125 GeV must be realized without resorting to heavy SUSY particles. 
Therefore, we impose the selection rule of the CP-even Higgs boson $h$ in our data as
\begin{equation}
123~\text{GeV}\leq m_h\leq 127~\text{GeV}.
\end{equation} 
If the other Higgs bosons are heavy, the Higgs sector will fall into the decoupling limit,
and the properties of $h$ will be SM-like which is preferred by the LHC data. 
Second, due to the null results of the SUSY searches at the LHC, several limits
 must be imposed on the masses of the colored particles, such as gluino and stop. 
So we will briefly summarize the current LHC bounds before discussing our results.

\subsection{Summary of Current LHC Bounds}

\begin{figure} [t]
\begin{center}
\includegraphics[width=0.49\textwidth]{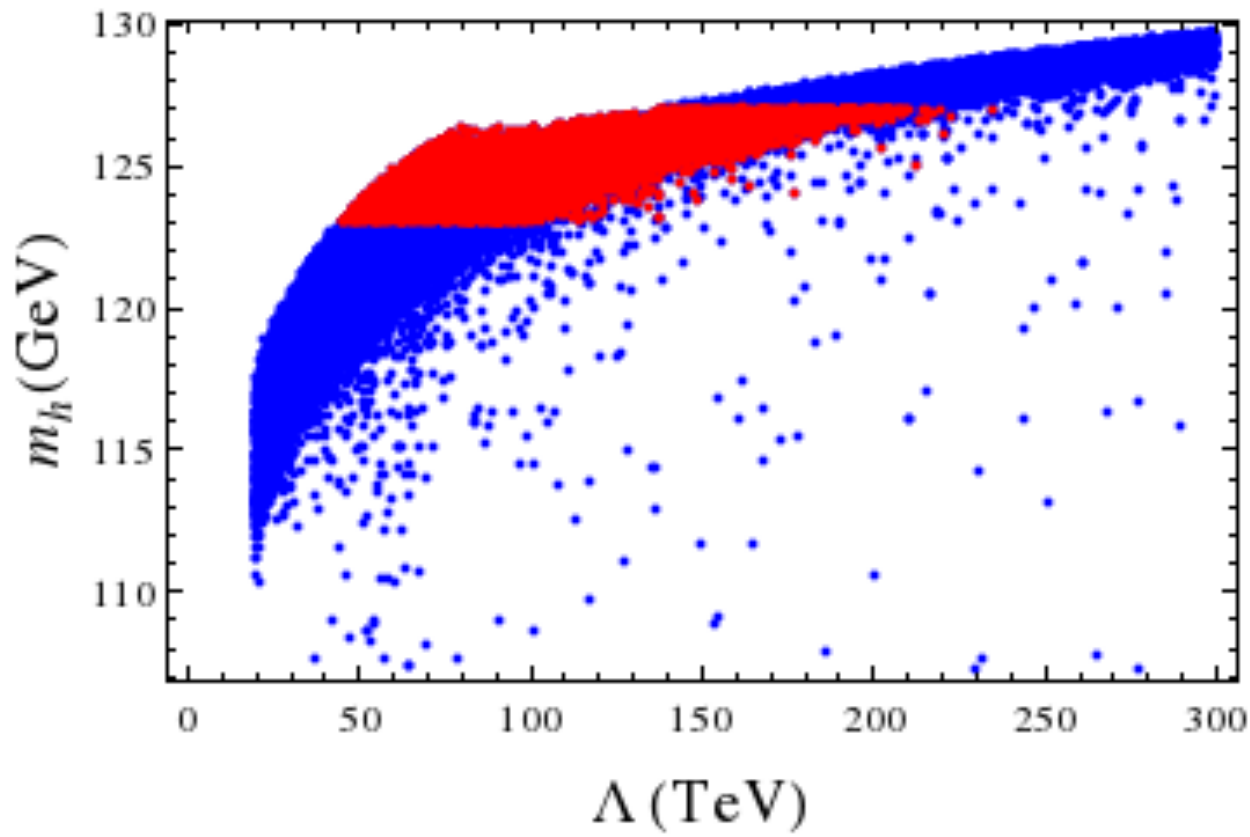}
\includegraphics[width=0.49\textwidth]{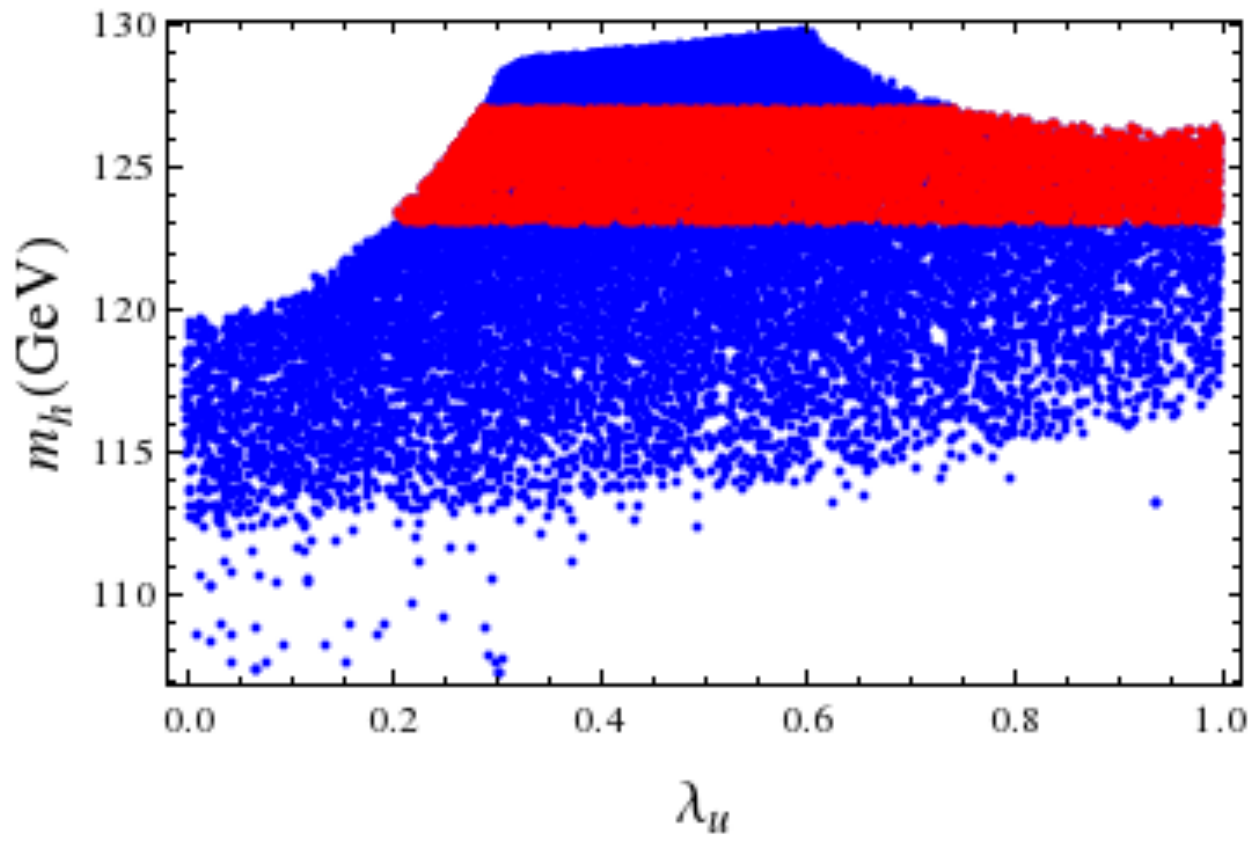}
\end{center}
\caption{(color online) The Higgs boson mass versus $\Lambda$ (left) and $\lambda_u$ (right). Blue points are whole scan results. 
Red points satisfy $123~\text{GeV}\leq m_h\leq 127~\text{GeV}$, $M_{\tilde{g}}\geq 1700~\text{GeV}$, $M_{\tilde{t}}\geq 700~\text{GeV}$, and $M_{\tilde{q}} \geq 800~\text{GeV}$. }
\label{fig:Higgs}
\end{figure}

This section is based on Ref.~\cite{Craig:2013cxa}. 
The current ATLAS and CMS summary plots can be found in Refs.~\cite{ATLAS_SUSY} and \cite{CMS_SUSY}, 
respectively. 
These plots present the sparticle mass low bounds for various SUSY search channels, 
which are based on the simplified models for the masses and branching ratios. 
For most of SUSY models, gluino is supposed to have large production cross-sections 
at the LHC due to the strong interaction. 
According to  Refs.~\cite{ATLAS_SUSY} and \cite{CMS_SUSY}, the strongest constraint on
 glunio mass comes from Ref.~\cite{Aad:2014wea}, where gluino is excluded for masses 
below 1700 GeV. The cascade decay of gluino is assumed to be 
$\tilde{g}\to\tilde{q}q$ and then $\tilde{q}\to q \tilde{\chi}^1_0$. The data, which focus 
on final states containing high-$p_T$ jets, missing transverse momentum, no electrons or muons, 
were recorded in 2012 by the ATLAS experiment in $\sqrt{s}$=8 TeV at the LHC with 
a total integrated luminosity of 20.3 $\text{fb}^{-1}$ \cite{Aad:2014wea}. 

The stop final state is also important because of the strong interaction as well as the 
relatively large Yukawa coupling. Before  the LHC, the light stop $\tilde t_1 $ in many natural 
SUSY scenarios is expected to have a mass below 1 TeV in order to avoid 
 a large fine-tuning. Depending on the mass assumptions, the following decay channels could be 
dominant: $\tilde t_1 \to t \tilde\chi_{1}^{0}$, $\tilde t_1 \to bW \tilde\chi_{1}^{0}$, $\tilde t_1 \to bff' \tilde\chi_{1}^{0}$ or $\tilde t_1 \to c \tilde\chi_{1}^{0}$ \cite{Aad:2012tx,Aad:2012yr,Aad:2013ija,Aad:2014qaa,Aad:2014mha,Aad:2014bva,Aad:2014kra,Aad:2014nra}. The searches are designed such that they cover all the possible decays of the stop into a neutralino LSP. For a massless $\tilde\chi_{1}^{0}$ the stop can be excluded up to 650-700 GeV (except some regions where the mass difference between the stop and the neutralino is near the top mass), while for $m_{\tilde\chi_{1}^{0}}>240$ GeV no limits can be provided. Limits on the first and second generation squark masses for simplified models are typically involved squark pair production $p p \to \tilde q \tilde q $ with only one decay chain $\tilde q \to q \tilde\chi^0_1$. Here it is assumed that the left and right-handed squarks has degenerate mass with gluino mass decoupled. As shown in Ref.~\cite{Aad:2014wea}, in this scenario squarks with a mass below about 800 GeV are excluded for a light neutralino.

In our model with a relatively large $\sqrt{F_X}$, the LSP is still gravitino.
Alhough all the SUSY particles will eventually decay into final states involving gravitino, 
these decays are extremely slow. The next lightest supersymmetric particle (NLSP) can be regarded as a stable particle at the collider scale and the gravitino will play no role in the collider physics. In our cases, the NLSP could be neutralino or stau depending on the parameter space. 
All above constraints are based on the assumption that heavy SUSY particles will decay into neutralino final state at the LHC.
If the NLSP is neutralino and stable at the collider scale, these constraints are still valid. 
If the NLSP is stau, the searches could be different as some stau final state might be recorded as charged tracks in the muon detector (for example, see \cite{Feng:1997zr,Hinchliffe:1998ys,Ambrosanio:2000ik,Ellis:2006vu}). 
In this paper, we naively impose the following selection rules of gluino mass and stop mass
\begin{align}
M_{\tilde{g}} &\geq 1700~\text{GeV},\\
M_{\tilde{t}} &\geq 700~\text{GeV},\\
M_{\tilde{q}} &\geq 800~\text{GeV},\,\,(\text{for the first and second generation squarks}).
\end{align}

\subsection{Particle Spectra and Fine-Tuning}

For simplicity, we fix the parameter $\tan\beta=10$. For all the other free parameters in our model, 
we do a random scan over them as below
\begin{align}
2\times 10^4~\text{GeV} &\leq \Lambda\leq 3\times 10^5~\text{GeV},\\
10^9~\text{GeV} &\leq M_{\text{mess}}\leq 10^{12}~\text{GeV},\\
0 &\leq\lambda_u \leq 1,\\
100~\text{GeV} &\leq\mu_0\leq 1000~\text{GeV}.
\label{eq:para4}
\end{align}
$\mu_0$ is given at the GUT scale. The RGEs are runnings from the GUT scale to the EW scale. 
At the messenger scale, the non-vanishing soft masses of the gauginos/sfermions and A-terms are generated as the threshold corrections in the RGEs. A successful EWSB is required, which will determine the exact values of $m^2_{H_u}(M_{\text{EW}})$ and $m^2_{H_d}(M_{\text{EW}})$. Based on the results at the EW scale, we will also run the RG evolutions back to the GUT scale to make sure $m^2_{H_u}(M_{\text{GUT}})>0$, $m^2_{H_d}(M_{\text{GUT}})>0$ and no Landau pole.

\begin{figure} [t]
\begin{center}
\includegraphics[width=0.49\textwidth]{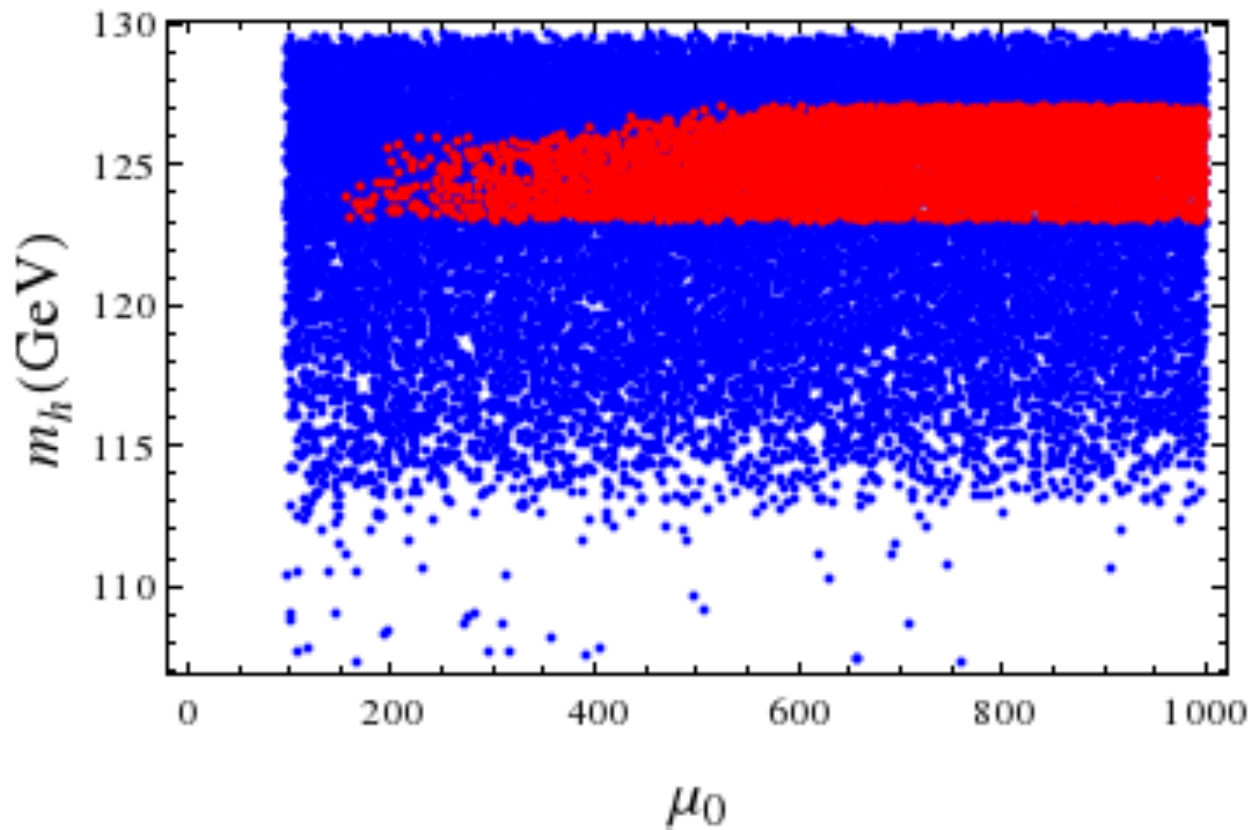}
\includegraphics[width=0.49\textwidth]{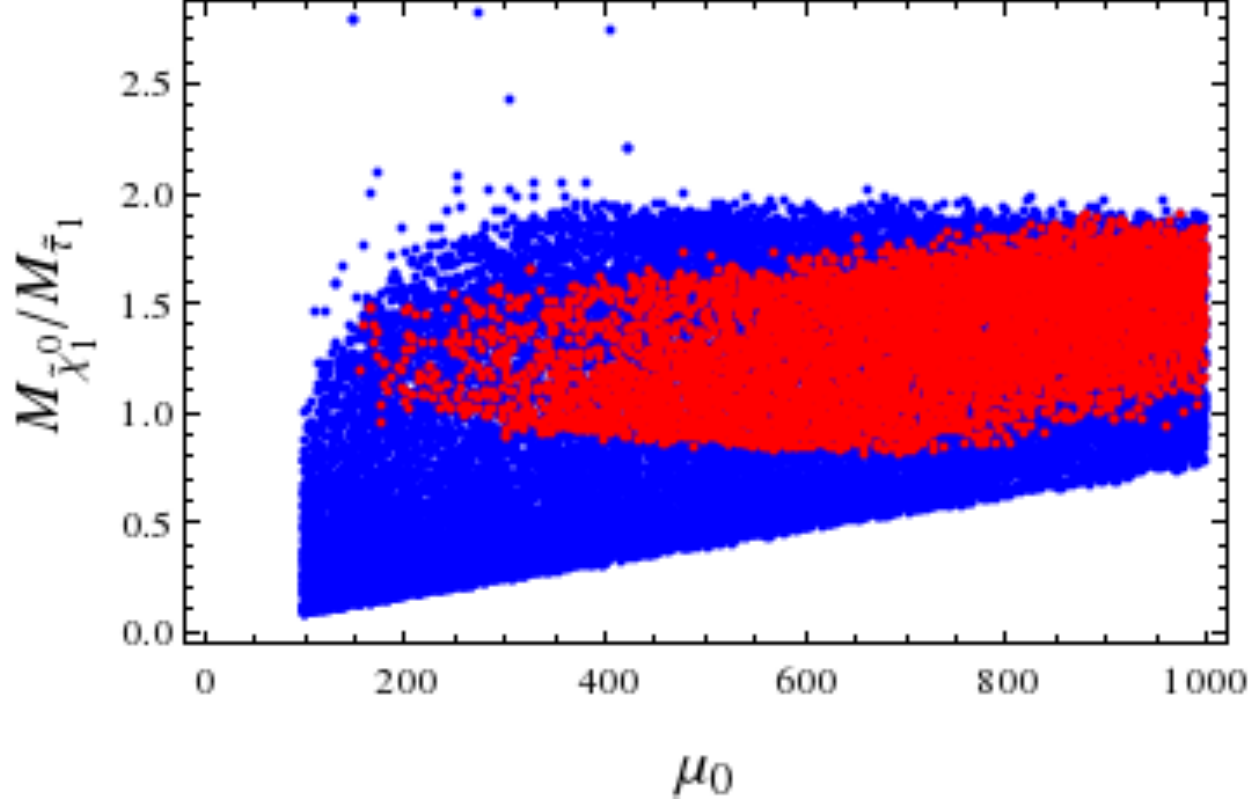}
\end{center}
\caption{(color online) $\mu_0$ dependence in our model. Blue points are corresponding to all scan results. Red points are corresponding to points with $123~\text{GeV}\leq m_h\leq 127~\text{GeV}$, 
$M_{\tilde{g}}\geq 1700~\text{GeV}$, $M_{\tilde{t}}\geq 700~\text{GeV}$, and $M_{\tilde{q}} \geq 800~\text{GeV}$. Left: Scan results shown in the [$\mu_0$, $m_h$] plane. Right: Scan results shown in the [$\mu_0$, $M_{\tilde{\chi}^0_1}/M_{\tilde{\tau}_1}$] plane. Here $\tilde{\chi}^0_1$ and $\tilde{\tau}_1$ are NLSP candidates in our model.}
\label{fig:Higgs2}
\end{figure}

First, we consider the light CP-even Higgs boson $h$ in our model. 
The distributions of its mass are given in Figs.~\ref{fig:Higgs} and \ref{fig:Higgs2}. 
Here, blue points are all the scan results, and
red points satisfy $123~\text{GeV}\leq m_h\leq 127~\text{GeV}$, 
$M_{\tilde{g}}\geq 1700~\text{GeV}$, $M_{\tilde{t}}\geq 700~\text{GeV}$, and $M_{\tilde{q}} \geq 800~\text{GeV}$, 
which are required by the LHC SUSY searches.
In the left panel of Fig.~\ref{fig:Higgs}, the Higgs mass $m_h$ is presented 
as a function of the parameter $\Lambda$. 
The mass window $123~\text{GeV}\leq m_h\leq 127~\text{GeV}$ is corresponding to a parameter window of $\Lambda$. 
The 125 GeV Higgs boson as well as relatively heavy gluino/stop prefer
 a relatively large $\Lambda$, because all the soft masses from gauge mediation are  proportional to it. 
A relatively heavy stop also significantly contributes to the Higgs mass $m_h$, 
as shown in Eq.~(\ref{eq:higgs mass}). 
In the right panel of Fig.~\ref{fig:Higgs}, we show how the Higgs mass $m_h$ depends 
on the parameter $\lambda_u$. All the red points with $123~\text{GeV}\leq m_h\leq 127~\text{GeV}$ are in the range 
with $\lambda_u>0.2$, where $\lambda_u$ will lead to a relatively large $A_t$ at the messenger scale. 
A relatively large $A_t$ plays a crucial role in lifting the Higgs boson mass to 125 GeV. A part of the parameter space with $\lambda_u>0.6$ has been excluded due to the requirement of a successful EWSB. Note that
$\lambda_u>0.6$ will result in a relatively large positive threshold contribution to
$\delta m^2_{H_u}$ at the messenger scale. 
When the RGEs run from the GUT scale down to the electroweak scale, $m^2_{H_u}$ fails to be negative 
due to such a large positive threshold effect $\delta m^2_{H_u}(M_\text{mess})$. Therefore, 
the EWSB can not be triggered in these cases. 
Our scan results in the [$\mu_0$, $m_h$] plane are shown in the left of Fig.~(\ref{fig:Higgs2}). 
One can see that the survived red points are almost independent of the parameter $\mu_0$ at the GUT scale. However, the GUT input $\mu_0$ will significantly influence the NLSP in our model. As the gravitino is the LSP, the lightest neutralino $\tilde{\chi}^0_1$ and the lightest stau $\tilde{\tau}_1$ are the NLSP candidates in our model. When $\mu_0$ is relatively small, NLSP in most cases is $\tilde{\chi}^0_1$ which is Higgsino-like, as shown in the right of Fig.~(\ref{fig:Higgs2}). When $\mu_0$ grows up, the Bino and Wino components of $\tilde{\chi}^0_1$ become important. 

\begin{figure} [t]
\begin{center}
\includegraphics[width=0.49\textwidth]{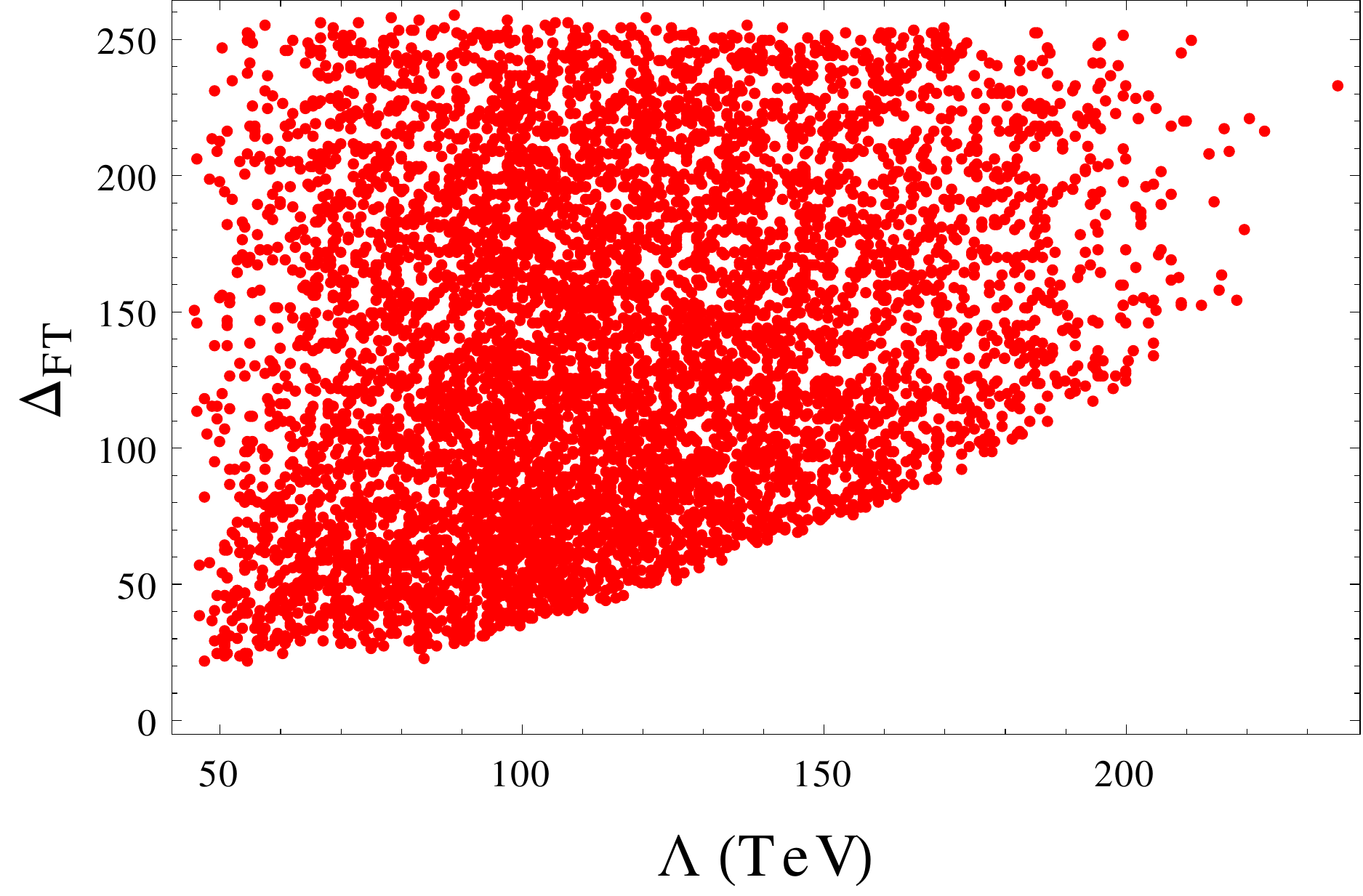}
\includegraphics[width=0.49\textwidth]{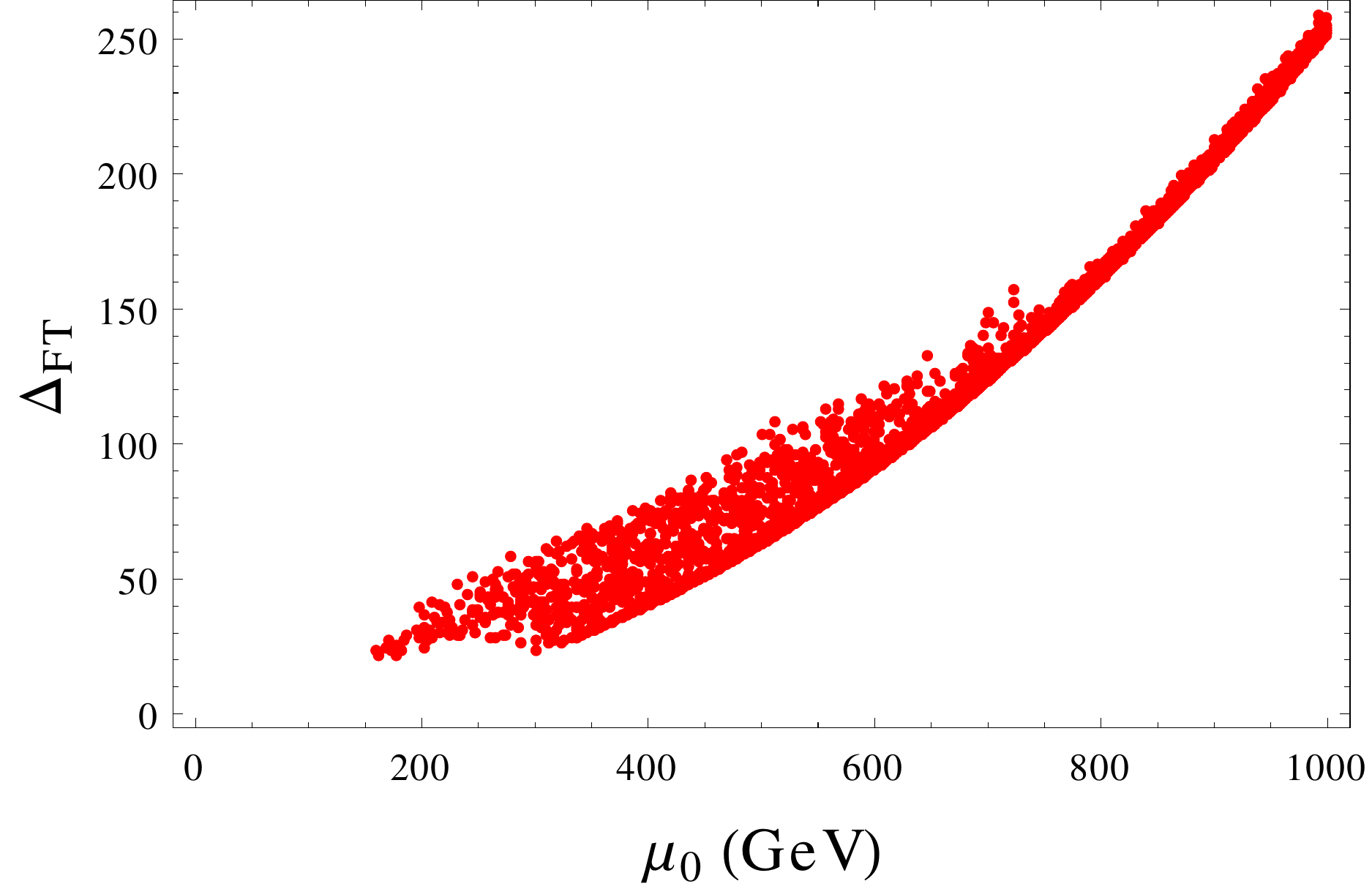}
\end{center}
\caption{ The fine-tuning measure $\Delta_{\text{FT}}$ versus $\Lambda$ (left) and $\mu_0$ (right) for all the red points.}
\label{fig:fine-tuning_random}
\end{figure}

For the survived red points which have $123~\text{GeV}\leq m_h\leq 127~\text{GeV}$, $M_{\tilde{g}}\geq 1700~\text{GeV}$, $M_{\tilde{t}}\geq 700~\text{GeV}$, and $M_{\tilde{q}}\geq 800~\text{GeV}$, 
we show their low-scale fine-tuning measure $\Delta_{\text{FT}}$ in Fig.~\ref{fig:fine-tuning_random}, 
which can be as low as 20 in our model. 
Obviously, the low-scale fine-tuning measure will become large if the GUT-scale
 input parameter $\mu_0$ grows up, 
which is shown in the right panel of Fig.~\ref{fig:fine-tuning_random}. 
This is because $\mu$-term is an important component for the definition of low-scale fine-tuning
 measure $\Delta_{\text{FT}}$, as shown in Eq.~(\ref{eq:FT}), and the
RGE runnings from $\Lambda_{\mathrm{GUT}}$ down to the electroweak scale only lead to 
a small correction to the $\mu$-term, {\it i.e.}, the low scale $\mu$-term is still 
dominated by its GUT-scale input $\mu_0$. 
For the ordinary GMSB models with new Yukawa couplings between 
the Higgs sector and messenger fields \cite{Kang:2012ra,Craig:2012xp,Albaid:2012qk,Byakti:2013ti,Evans:2013kxa,Knapen:2013zla,Ding:2013pya,Ding:2014bqa}, 
the large A-terms as well as a positive soft mass $m^2_{H_u}$ are generated at the messenger scale.
Compared to $5 \oplus \bar5$ models, such positive soft mass $m^2_{H_u}$ at $M_{\text{mess}}$ is small in our $10 \oplus \bar10$ models due to the negative contribution from $g_3$. Moreover, it is easier in our model to obtain a negative $m^2_{H_u}$ at the electroweak scale because our boundary condition of $m^2_{H_u}$ is given at the GUT scale.
When the RGEs run from the GUT scale to messenger scale, the Yukawa coupling $Y_t$ will persistently 
provide negative contributions to $m^2_{H_u}$ 
even if all the gaugino masses are still vanishing during the running. 
The EWSB is guaranteed for the survived points. In addition, for large $\lambda_u$ $(>0.6)$, the negative 
contributions to $m_{Q_3}^2$ and $m_{u_3}^2$ become comparable with the trilinear $A_t$ term
 and reduce the stop masses significantly.  As a consequnence, the Higgs boson mass is reduced 
at large $\lambda_u$, which can be found in the right panel of Fig.~\ref{fig:Higgs}.

\begin{figure} [t]
\begin{center}
\includegraphics[width=0.48\textwidth]{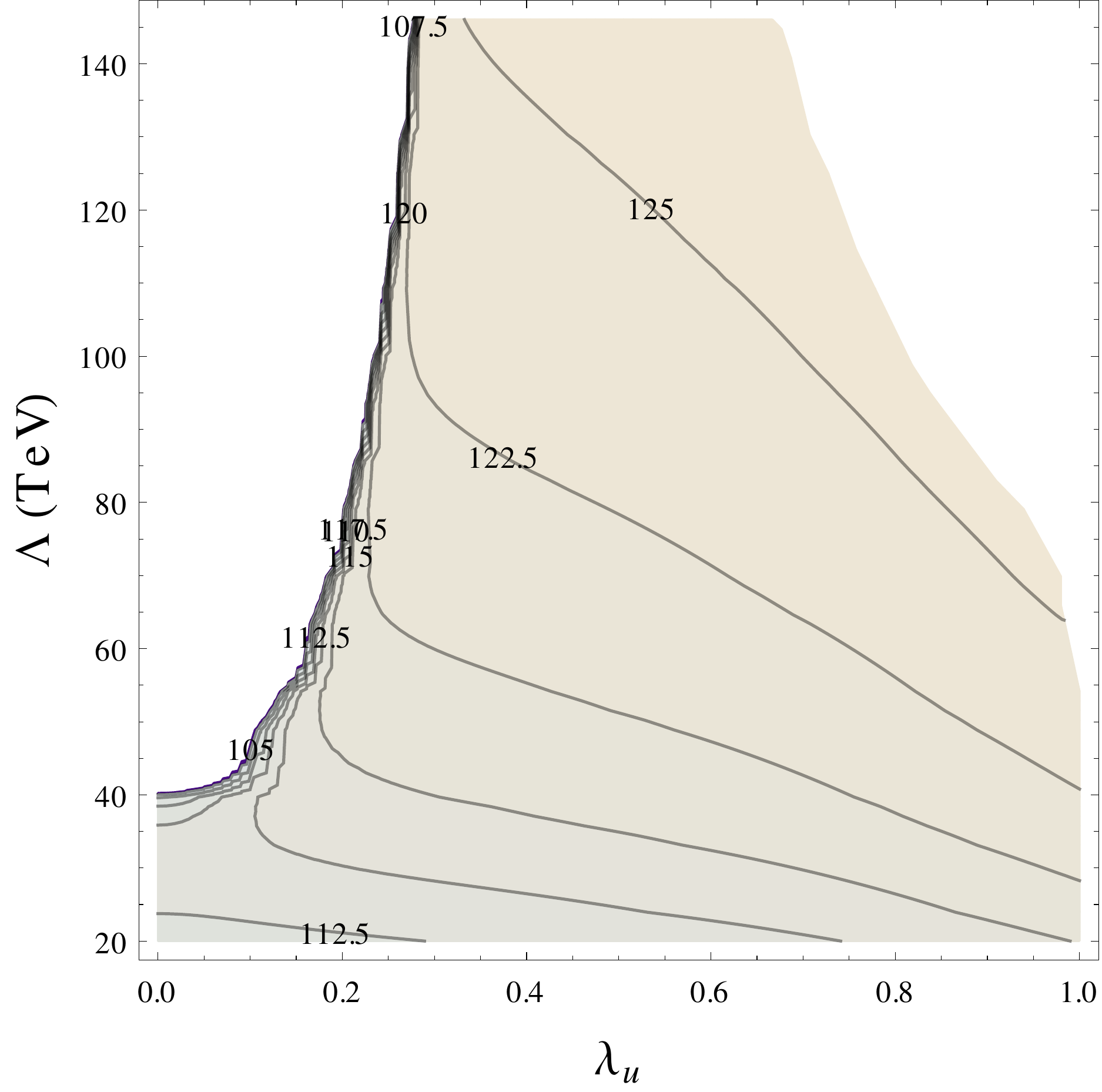}
\includegraphics[width=0.48\textwidth]{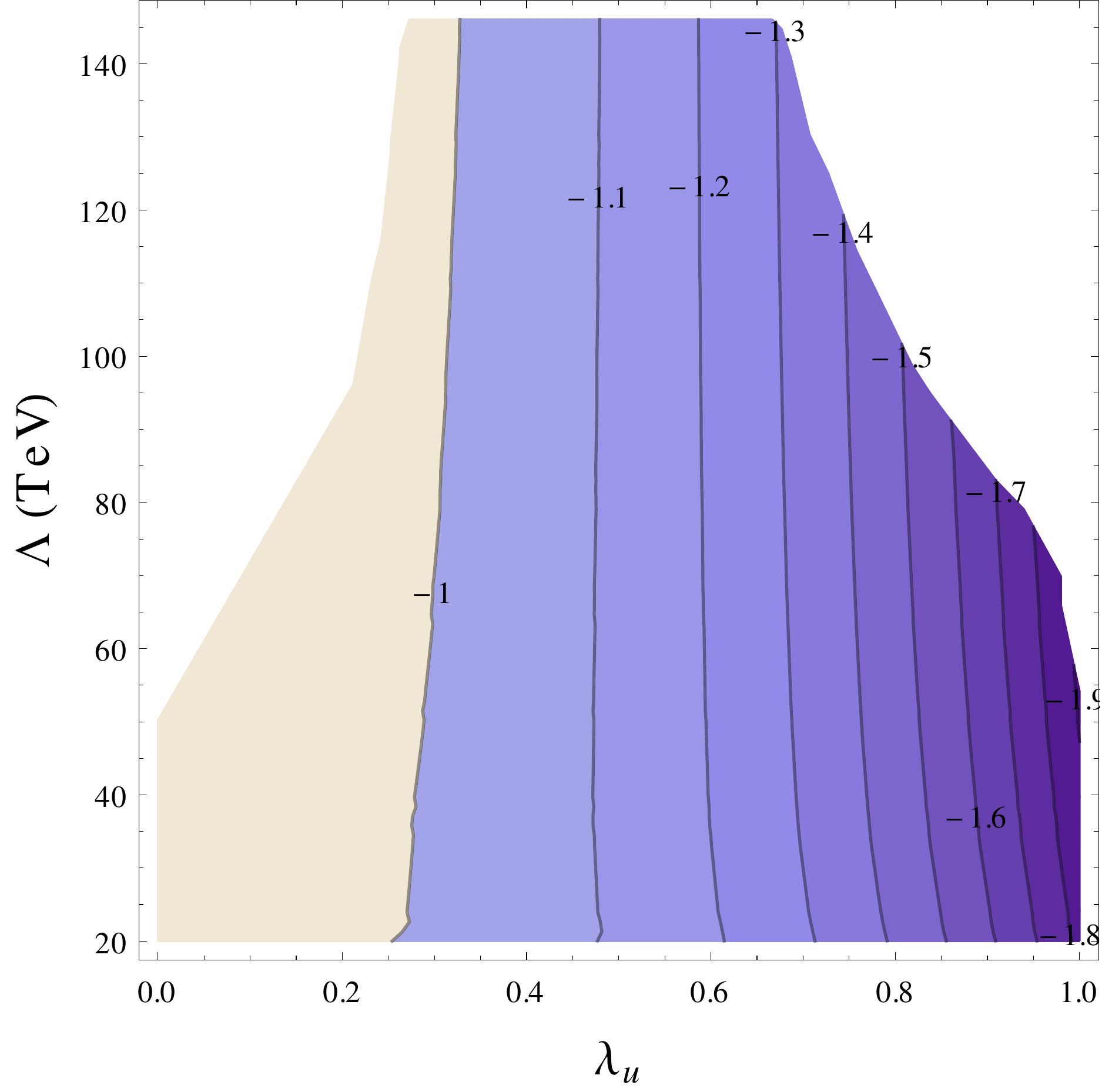}
\end{center}
\caption{The contour plots of $m_h$ (left) and $\tilde{A_t}/M_{\mathrm{SUSY}}$ (right) in the [$\lambda_u$, $\Lambda$] planes.}
\label{fig:Higgs3}
\end{figure}

\begin{figure} [t]
\begin{center}
\includegraphics[width=0.48\textwidth]{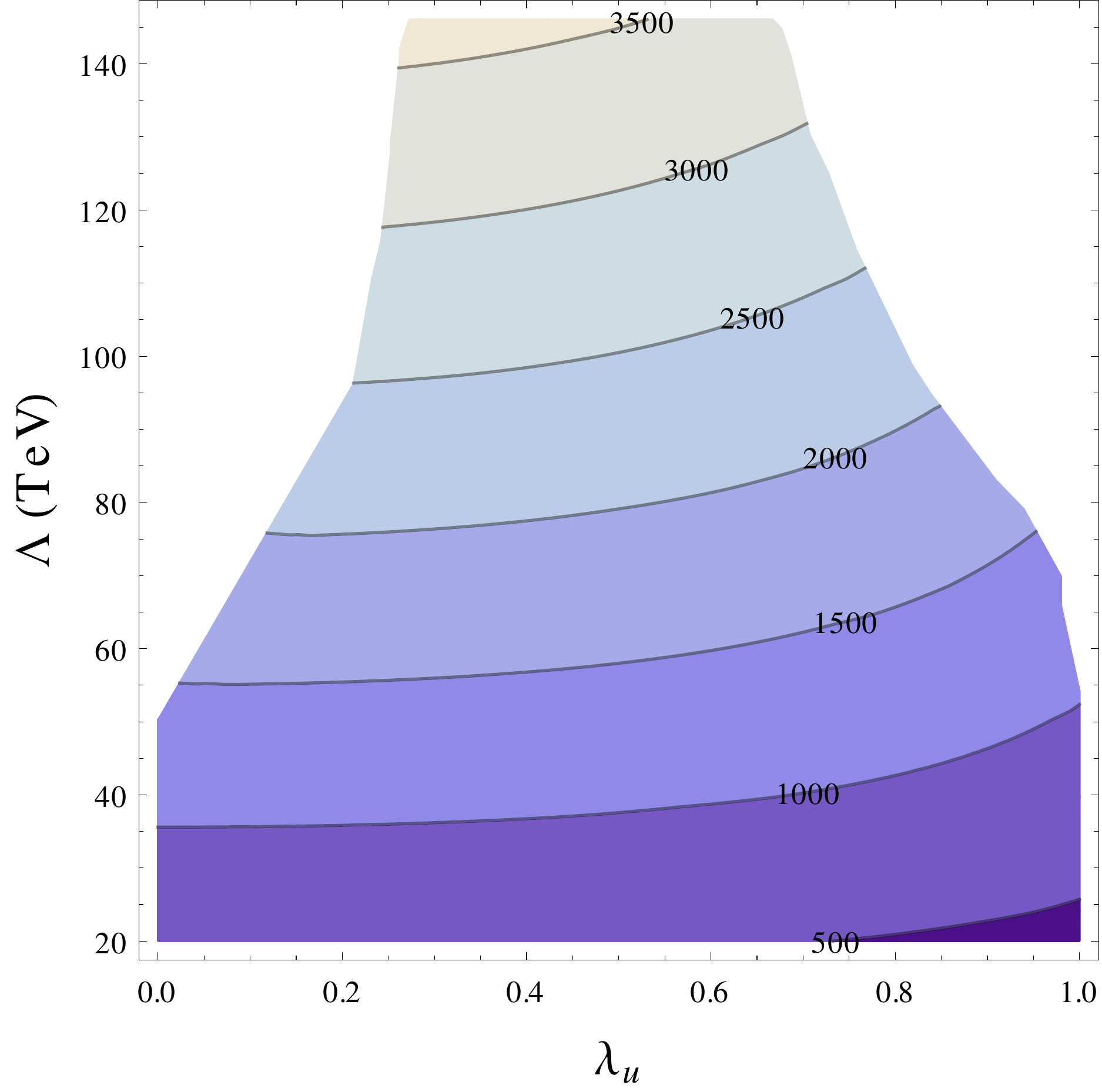}
\includegraphics[width=0.48\textwidth]{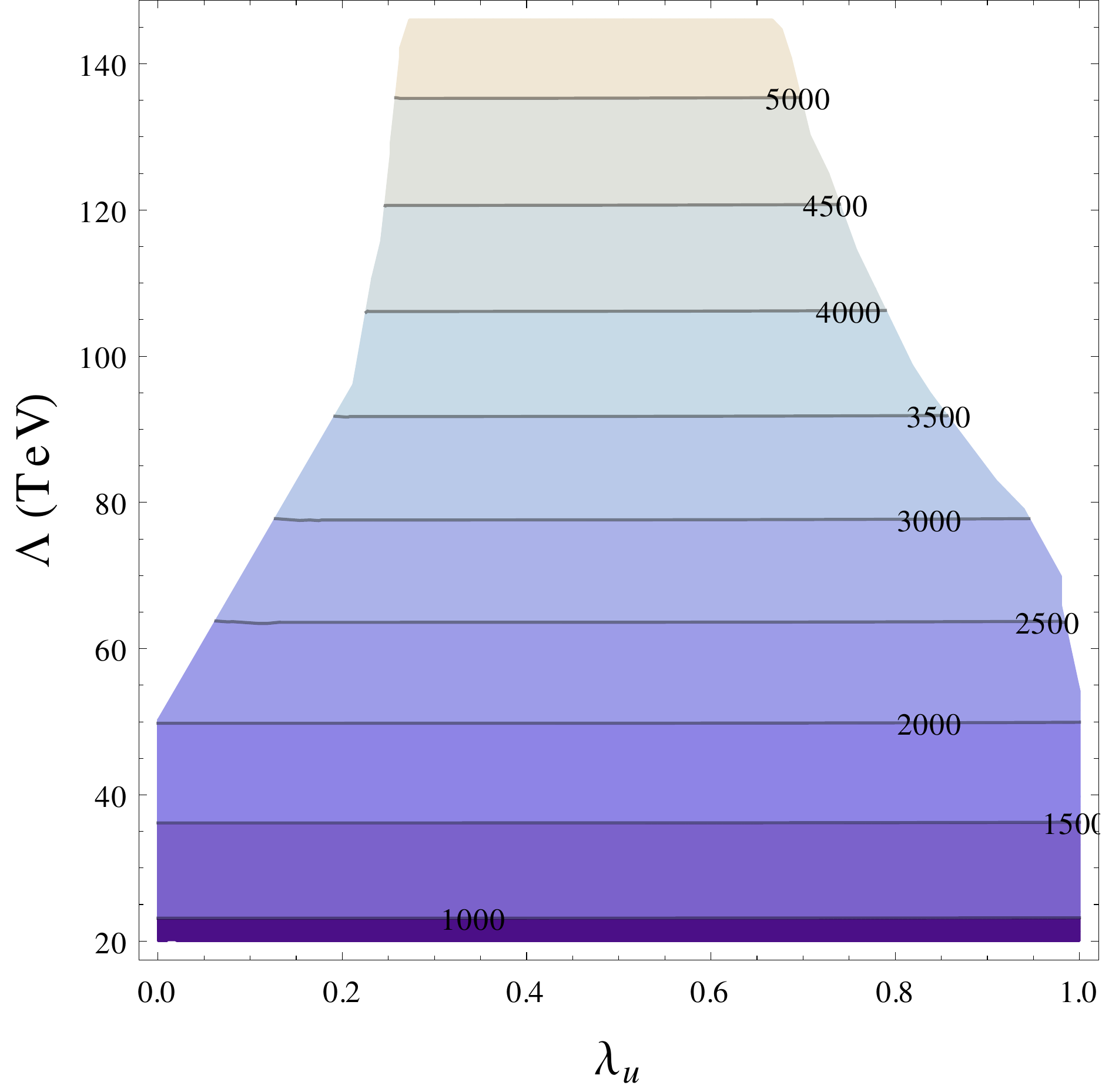}
\end{center}
\caption{The contour plots of the masses of the light stop (left) and gluino (right) in the [$\lambda_u$, $\Lambda$] planes.}
\label{fig:gluino_stop}
\end{figure}

We would like to focus on the survived red points and study more features about them. 
Therefore, we make a careful scan for $M_{\text{mess}}=10^{10}$ GeV and $\mu_0=150$ GeV. 
About the other free parameters in our model, we choose
\begin{align}
2\times10^4~\text{GeV} &\leq \Lambda\leq 3\times 10^5~\text{GeV},\\
0 &\leq\lambda_u \leq 1.
\label{eq:para4}
\end{align}
The contour plots of $m_h$ and $\tilde{A_t}/M_{\mathrm{SUSY}}$ in the $\Lambda$ versus $\lambda_u$ planes
are shown in the left and right panels of Fig.~\ref{fig:Higgs3}, respectively.
A Higgs boson with mass around 125 GeV is corresponding to the region $\tilde{A_t}/M_{\mathrm{SUSY}}>-1$. 
Such relatively large $A_t$-terms are generated by a relatively large coupling $\lambda_u$ 
between the Higgs and the messenger fields.
If we turn off the coupling $\lambda_u$, the Higgs mass $m_h$ will be smaller than 116 GeV which is 
excluded by the current LHC results.
With a relatively large $A_t$-term, the masses of the light stop and gluino are shown in Fig.~\ref{fig:gluino_stop}. 
The light stop can be as light as 700 GeV in our scenario while the gluino can be lighter than 1.8 TeV, 
both of which are quite different from the exact sweet spot SUSY, where
both stop and gluino should be heavier than 5 TeV 
in order to obtain a $125$ GeV Higgs boson \cite{Fukushima:2013vxa}. 
Thus, adding the extra Higgs-messenger coupling $\lambda_u$ is an solution to the heavy spectrum problem. 
For such light SUSY particle spectra in our model, this scenario can definitely be tested by the run II of the LHC. 
Moreover, the naturalness condition is kept due to the light SUSY particle spectra, and
there is no heavy flavor problem as the soft masses of gauginos/sfermions are all based on gauge mediation.

\begin{figure}
\begin{center}
\includegraphics[width=0.42\textwidth]{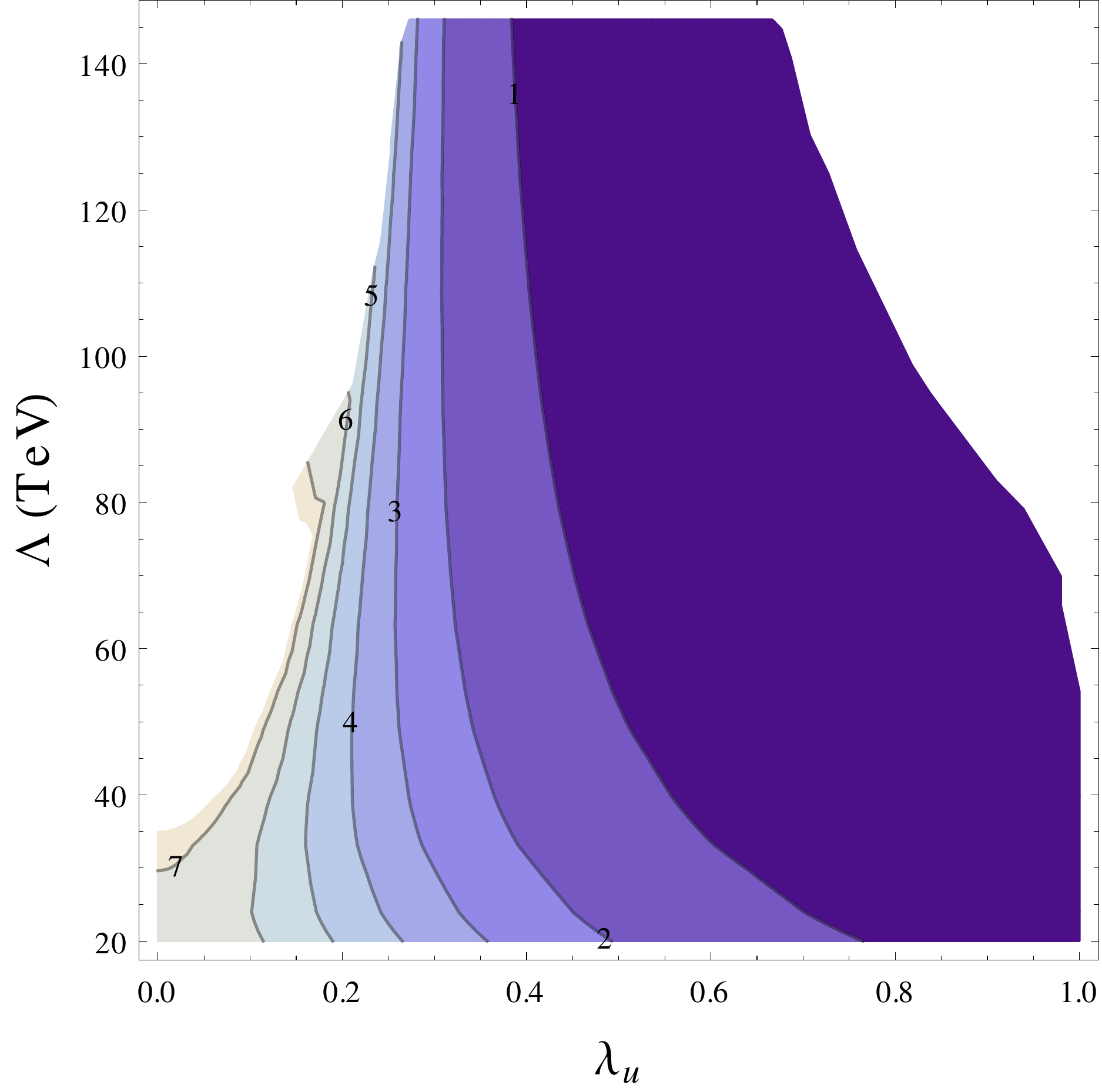}
\includegraphics[width=0.485\textwidth]{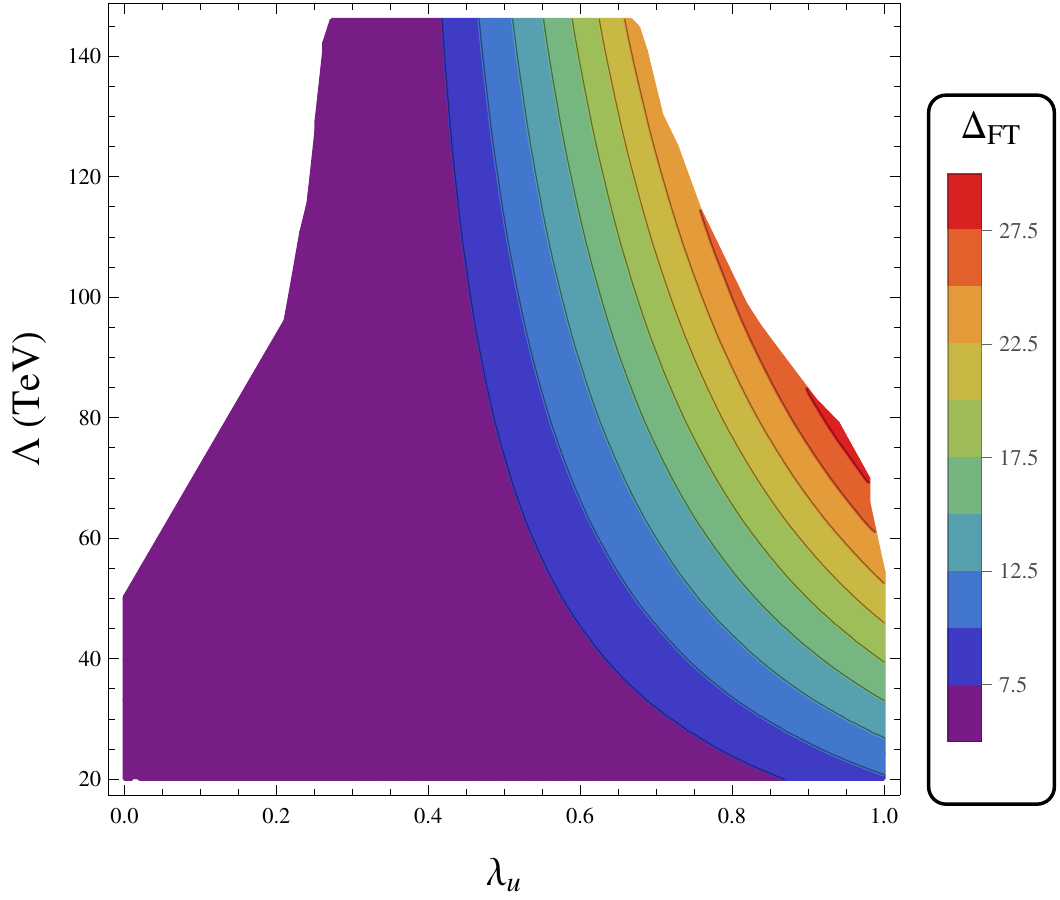}
\end{center}
\caption{The contour plots of the ratio $\mu^2/B_{\mu}$ at the electroweak scale (left) and the low-scale fine-tuning measure $\Delta_{\text{FT}}$ (right) in the [$\lambda_u$, $\Lambda$] planes.}
\label{fig:mu}
\end{figure}

Compared to the ordinary GMSB, the framework of sweet spot SUSY provides a solution 
to the $\mu$-$B_{\mu}$ problem. 
In the left panel of Fig.~\ref{fig:mu}, we show the ratio of $\mu^2/B_{\mu}$ at the electroweak scale.
A Higgs boson with mass around 125 GeV is corresponding to the region with of $\mu^2/B_{\mu}\sim\mathcal{O}(1)$. 
Here, the $\mu$-term is generated at the GUT scale from the direct coupling between the hidden sector 
and Higgs sector, and
the RGE correction to $\mu$-term is tiny from the GUT scale down to electroweak scale. 
For the $B_{\mu}$-term, we have $B_{\mu}=0$ at the GUT scale due to the approximate PQ symmetry,
and a non-vanishing $B_{\mu}$-term at the electroweak scale is obtained by the RGE runnings. 
In addition, when $\lambda_u$ grows up, we see from the left panel of Fig.~\ref{fig:mu} 
that the $B_{\mu}$-term at the electroweak scale increases as well. 
In the right panel of Fig.~\ref{fig:mu}, we show the distribution of the low-scale fine tuning 
measure $\Delta_{\text{FT}}$ in the [$\lambda_u$, $\Lambda$] plane. 
A Higgs boson with mass around 125 GeV can be corresponding to the region where $\Delta_{\text{FT}}$ is as low as 20. 
$\Delta_{\text{FT}}$ increases if $\lambda_u$ grows up in the region with a 125 GeV Higgs boson. 
This is because $\Delta_{\text{FT}}$ is dominated by the $B_{\mu}$-term in this region 
since we fix the input parameters $\mu_0$ and $\tan\beta$ in this careful scan.

\begin{figure}
\begin{center}
\includegraphics[width=0.48\textwidth]{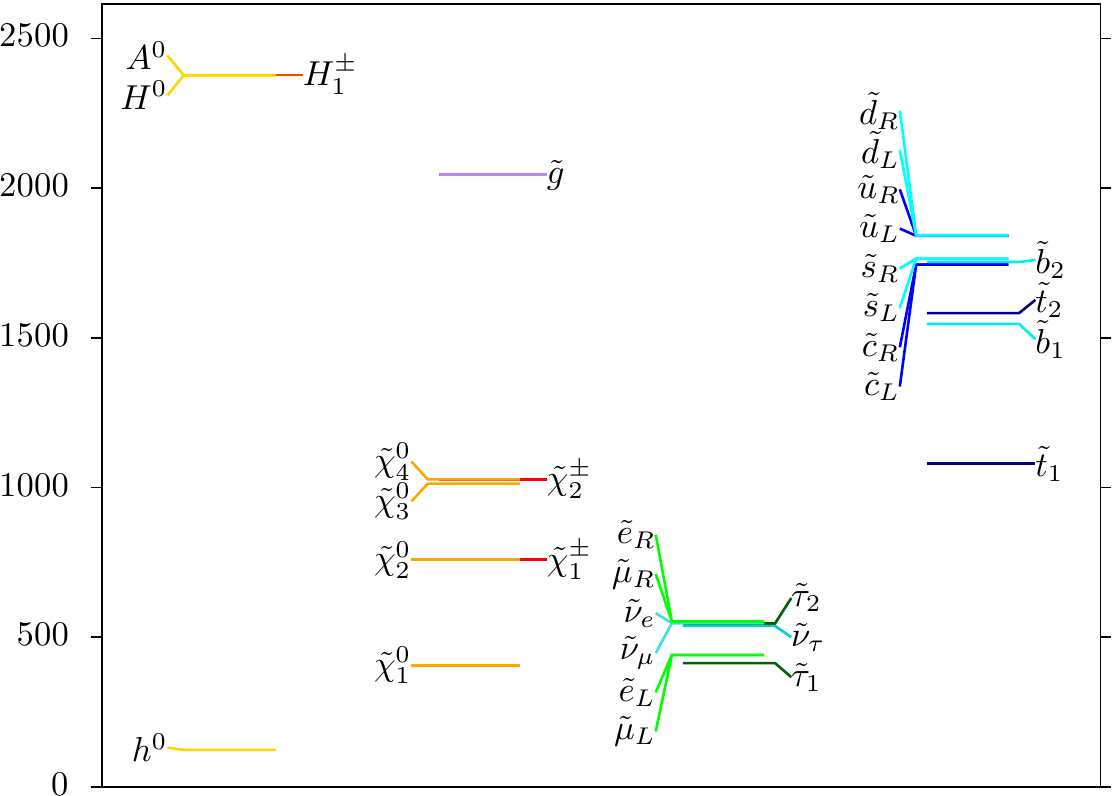}
\includegraphics[width=0.48\textwidth]{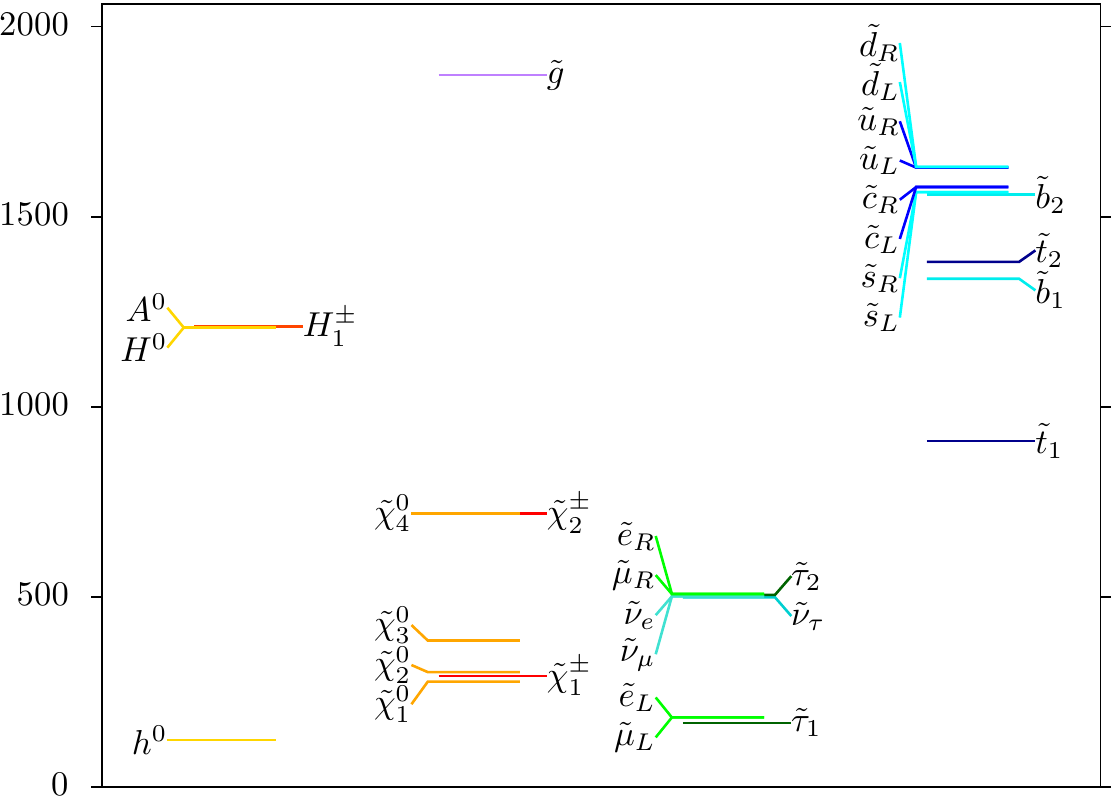}
\end{center}
\caption{Two benchmark points in our model with neutralino NLSP (left) and stau NLSP (right).}
\label{fig:benchmark}
\end{figure}

In a summary, we present an interesting SUSY scenario which is theoretically interesting and 
simply predicted by only five free parameters. In particular,
the 125 GeV Higgs boson can be realized naturally, and there are no flavor problem and $\mu$-$B_{\mu}$ problem. In Fig.~(\ref{fig:benchmark}), we list the spectra of two benchmark points in our model. In the left panel, the lightest neutralino is the NSLP candidate. In the right panel, the NSLP candidate is the lightest stau. In both cases, glunio and stop are relatively light, which can be tested at the upcoming run II of the LHC experiment. A thorough analysis of searching these scenarios at the LHC will be performed in a future publication. 

\subsection{Gravitino Dark Matter}

Gravitino is the LSP in our model. The gravitino mass should not be larger than $\mathcal{O}(1)$ GeV, otherwise, the flavor problem 
will be generated due to gravity mediation. 
Interestingly, such a gravitino dark matter can come from a thermal production and be consistent 
with the thermal leptongenesis. 
The baryon number asymmetry $\Omega_b$ can be produced by thermal leptogenesis, which is given by
\begin{equation}
\Omega_b\leq 0.04\left(\frac{T_R}{10^9\,\text{GeV}}\right),
\end{equation}
with $T_R$ being the reheating temperature. 
In order to realize the observed value $\Omega_b=0.0499$ \cite{Ade:2013zuv}, 
one has $T_R\geq 10^9\,\text{GeV}$ \cite{Davidson:2002qv,Giudice:2003jh,Buchmuller:2004nz,Buchmuller:2005eh}. 
In the  thermal leptogenesis, it is difficult to realize the observed value 
$\Omega_{\text{dm}}=0.265$ \cite{Ade:2013zuv} if gravitino is the dark matter candidate. 
This is because the relic abundance of thermally produced gravitino is usually also proportional to $T_R$ \cite{Fukugita:1986hr,Bolz:1998ek,Bolz:2000fu,Pradler:2006hh}. Under these conditions, the correct 
ratio $\Omega_{\text{dm}}/\Omega_{b}\sim 5$ can not be realized.

However, the estimation of the relic abundance for thermally produced gravitino should be corrected. 
The relic density is still fixed by $T_R$ if $T_R<M_{\text{mess}}$, but it can be insensitive to 
the reheating temperature if $T_R>M_{\text{mess}}$ \cite{Choi:1999xm,Fukushima:2013vxa}. 
For $T_R>M_{\text{mess}}$, the relic density is  \cite{Fukushima:2013vxa}
\begin{equation}
\Omega_{3/2}~h^2\simeq 370\left(\frac{M_{\text{mess}}}{10^6\,\text{GeV}}\right)\left(\frac{\text{GeV}}{m_{3/2}}\right)\left(\frac{m_{\tilde{g}}}{5~\text{TeV}}\right)^2
+0.53\left(\frac{T_R}{10^{13}~\text{GeV}}\right)\left(\frac{m_{3/2}}{\text{GeV}}\right).\label{eq:relic}
\end{equation}
The former contribution in the right-handed side of Eq.~(\ref{eq:relic}) comes from the longitudinal mode 
of the gravitino, while the latter arises from the transverse component. 
When the reheating temperature is higher than messenger scale, the thermally produced gravitino 
and thermal leptogenesis can be compatible so that the observed ratio $\Omega_{3/2}/\Omega_b=5$ can be realized. 
In order to get the correct values $\Omega_{\text{dm}}=0.265$ and $\Omega_b=0.0499$, a late-time entropy release 
is required.
The SUSY breaking field $X$ can be the pseudo-modulus field which provides an appropriate dilution 
factor~\cite{Fukushima:2013vxa}. 
Compared to the exact sweet spot SUSY discussed in Refs.~\cite{Ibe:2007km,Ibe:2007gf,Ibe:2007mr,Fukushima:2013vxa}, 
our modified model can predict a relatively light spectra which can be checked by the run II of the LHC. 
In the mean time, there still exists large viable parameter space to account 
for the cosmological observations. 
The thermal production of gravitino as well as thermal leptogenesis can still be realized, 
and the discussion should be similar to that in Ref.~\cite{Fukushima:2013vxa}. 

\section{Conclusion}
\label{sec:conclusion}

The discovery of a 125 GeV SM-like Higgs boson as well as the natural SUSY assumption suggest a large $A_t$ term 
in the MSSM. So in the GMSB, the extended Higgs-messenger coupling is always introduced
 to generate the non-vanishing $A$-terms at the messenger scale. 
However, the $\mu$-$B_{\mu}$ problem is still unsolved unless one considers the NMSSM. 
Since the run II of the LHC will start soon, it is important to think about the 
feasible SUSY models which describe new physics at the TeV scale and can be detected by the coming LHC experiments. 
In this paper, we have proposed the MSSM with the GMSB, Higgs-messenger interaction,
and generalized GM mechanism.  At the GUT scale,
the SUSY breaking sector and Higgs fields are assumed to be directly coupled.
Because of the approximate PQ symmetry, only the $\mu$-term and soft masses $m_{H_u}$/$m_{H_d}$ are 
generated. While
the sfermion soft masses, gaugino masses, $A$-terms, and $B_{\mu}$-term are all vanished.
Below the GUT scale, it is effectively the GMSB with extended Higgs-messenger coupling. 
The RGEs are run from the GUT scale down to EW scale. 
At the messenger scale the messenger fields are integrated out.
The non-vanishing soft masses of the gauginos/sfermions and A-terms are generated as the threshold corrections 
in the RGE runnings. Especially,
a large non-vanishing $A_t$-term at the messenger scale is produced by the extended Higgs-messenger coupling. 
So our model can have a SM-like Higgs boson at 125 GeV without moving forward into the split SUSY. 
In addition, it is easier in our model to obtain a negative $m^2_{H_u}$ at the EW scale because
 our boundary condition of $m^2_{H_u}$ is given at the GUT scale.
When the RGEs run from the GUT scale to messenger scale, the Yukawa coupling $Y_t$ will persistently 
provide a negative contributions to $m^2_{H_u}$. The EWSB is guaranteed in our model as we run the RGE of $m^2_{H_u}$ for a long energy scale range
 from the GUT scale. 
On the theoretical aspect, gauge coupling unification is guaranteed. The flavor problem and $\mu$-$B_{\mu}$ problem are solved.
On the phenomenological aspects, our model has only five free parameters,
 can predict a 125 GeV SM-like Higgs boson, and evades all the current LHC SUSY search constraints. 
The low-scale fine-tuning measure can be as low as 20 with the light stop mass below 1 TeV and 
gluino mass below 2 TeV. 
Since glunio and stop can be relatively light, this natural SUSY model
 could be tested at the upcoming run II of the LHC experiment.

Furthermore, the gravitino mass $m_{3/2}$ is typically smaller than $\mathcal{O}(1)$ GeV in order to evade 
the flavor constraints. 
Due to a relatively large $\sqrt{F_X}$, the gravitino will play no role in the collider physics.
Interestingly, the gravitino can be a good dark matter candidate. Such a gravitino dark matter can 
come from a thermal production with the correct relic density and be consistent with the thermal leptongenesis.

\section*{Acknowledgements}
We would like to thank Qaisar Shafi and Florian Staub for very useful discussion. T.L. is supported in part
by the Natural Science Foundation of China under grant numbers 10821504, 11075194, 11135003, 11275246, 
and 11475238, and by the National Basic Research Program of China (973 Program) under grant number 2010CB833000.
L.W. is supported by the DOE Grant No.DE-FG02-12ER41808.

\bibliographystyle{JHEP}
\bibliography{SweetSUSY}
\end{document}